\documentclass[12pt]{article}
 \usepackage{amssymb,color,graphicx,epsfig,epsf}
    \usepackage{latexsym,amsmath}

\newcommand{\Or}{{\cal O}}
\newcommand{\bc}{\begin{center}}
\newcommand{\ec}{\end{center}}
\def\ba#1{\begin{array}{#1}\displaystyle}
\newcommand{\ea}{\end{array}}
\newcommand{\z}{\\[2mm] \displaystyle}

\newcommand{\beq}{\begin{equation}}
\newcommand{\eeq}{\end{equation}}
\newcommand{\beqa}{\begin{eqnarray}}
\newcommand{\eeqa}{\end{eqnarray}}
\newcommand{\no}{\nonumber}
\newcommand{\n}{\nonumber\\}
\newcommand{\bi}{\begin{itemize}}
\newcommand{\ei}{\end{itemize}}
\def\mato#1{\left(\ba{#1}} % exemple: \mato{cc} a & b \\ c & d \matf
\def\matf{\ea\right)}

\def\lt#1{\left#1}
\def\rt#1{\right#1}
\def\t#1{\tilde{#1}}
\def\h#1{\hat{#1}}
\def\b#1{\bar{#1}}
\def\frc#1#2{\frac{#1}{#2}}

\newcommand{\p}{\partial}

\newcommand{\vac}{{\rm vac}}
\newcommand{\bra}{\langle}
\newcommand{\ket}{\rangle}
\newcommand{\Z}{{\mathbb{Z}}}

\newcommand{\R}{{\mathbb{R}}}

\newcommand{\ep}{\epsilon}
\newcommand{\varep}{\varepsilon}

\newcommand{\Tr}{{\rm Tr}}

\newcommand{\ft}{{\cal L}}
\newcommand{\braL}{\langle}
\newcommand{\ketL}{\rangle_T}
\newcommand{\vacft}{{\rm vac}_{\cal L}}

\newcommand{\zt}{{\cal U}}
\newcommand{\shift}{{\rm v}}
\newcommand{\sym}{{\rm sym}}

\newcommand{\dd}{{\rm d}}
\newcommand{\xx}{{s}}

\begin{document}

\pagestyle{empty} \pagenumbering{arabic}

\vspace{2cm} \bc \Large\bf  Integral equations and large-time asymptotics\\ for finite-temperature\\ Ising chain correlation functions \ec
\vspace{1cm} \bc {\large Benjamin Doyon}

{\em Department of Mathematical Sciences\\
Durham University\\
South Road, Durham DH1 3LE, U.K.}
\ec
\bc {\large Adam Gamsa}

{\em Rudolf Peierls Centre for Theoretical Physics\\
University of Oxford\\
1 Keble Road, Oxford OX1 3NP, U.K.} \ec
\vspace{1cm} \bc \bf
Abstract \ec
This work concerns the dynamical two-point spin correlation functions of the transverse Ising quantum chain at finite (non-zero) temperature, in the universal region near the quantum critical point. They are correlation functions of twist fields in the massive Majorana fermion quantum field theory. At finite temperature, these are known to satisfy a set of integrable partial differential equations, including the sinh-Gordon equation. We apply the classical inverse scattering method to study them, finding that the ``initial scattering data'' corresponding to the correlation functions are simply related to the one-particle finite-temperature form factors calculated recently by one of the authors. The set of linear integral equations (Gelfand-Levitan-Marchenko equations) associated to the inverse scattering problem then gives, in principle, the two-point functions at all space and time separations, and all temperatures. From them, we evaluate the large-time asymptotic expansion ``near the light cone'', in the region where the difference between the space and time separations is of the order of the correlation length.
\vfill\noindent November 2007

\newpage
\pagestyle{plain} \setcounter{page}{1}

%%%%%%%%%%%%%%%%%%%%%%%%%%%%%%%%%%%%%%%%%%%%%%%%%%%%%%%%%%
\section{Introduction}
%%%%%%%%%%%%%%%%%%%%%%%%%%%%%%%%%%%%%%%%%%%%%%%%%%%%%%%%%%
The Ising model has a long and illustrious history, due to its simplicity of formulation, its integrability and its applicability to physical systems. We shall be considering the quantum model in one spatial plus one temporal dimension, a spin-half chain which describes the behaviour of strongly anisotropic systems such as ${\rm Dy}({\rm C}_2{\rm H}_5 {\rm SO}_4)_39{\rm H}_2{\rm O}$, ${\rm FeCl}_22{\rm H}_2O$ and related compounds~\cite{Bikas}. The Hamiltonian is written as below, with $\sigma_{i}^{x,z}$ representing the Pauli spin matrices at site $i$
\begin{equation}\label{spinchain}
H_{I}=-J\sum_{i}\sigma_{i}^{z}\sigma^{z}_{i+1}+g\sum_{i}\sigma_i^x~,
\end{equation}
where $g$ describes the strength of an externally applied magnetic field. For fixed $J$, the parameter $g$ controls the ground state of the system, with $g<J$ favouring a state with spins aligned (spin up) or antialigned (spin down) along the $z$ axis and  $g>J$ leading to a ground state with spins aligned along the direction of the magnetic field. At intermediate values, the magnetic field induces tunelling between the spin up and down states. There is a duality mapping from the $g>J$ regime to the same model with $g<J$. The self dual value of the couplings, $g_c=J$, is a quantum critical point. In the neighbourhood of this point, the low-energy, universal behaviour of the model is described simply by the quantum field theory of free Majorana fermions with $m\propto g_c-g$ (in conformal field theory language, it is the Ising minimal model deformed by the energy operator). We will study this region; it occurs around any quantum critical point in the Ising universality class, hence is more general than the quantum Ising model itself.

The quantities that are of most interest and usually accessible experimentally in spin chains are the expectation values at temperature $T$ of products of time-evolved spin matrices, such as $\Tr\lt(e^{-H_I/T} \sigma^{z}_{i}(t)\sigma^{z}_{j}(0)\rt) / \Tr\lt(e^{-H_I/T}\rt)$, on which we will concentrate in the present paper. These may be related to response functions to magnetic perturbations which can be measured by neutron scattering experiments, for example. The scaling limit of these quantities near the quantum critical point is obtained by taking $|i-j| \propto J/|g-g_c|$, $t\propto 1/|g-g_c|$ and $T\propto |g-g_c|$ as $g\to g_c$ from above (disordered regime) or below (ordered regime).

In the scaling limit, these quantities are related in a non-trivial fashion to the expectation values of products of fields in the fermionic model, due to the non-local nature of the mapping to the field theory. They are correlation functions of twist fields related to the $\Z_2$ symmetry of the fermions, as recalled in section~\ref{SecOrdDis}. Despite considerable effort, no systematic method exists for determining the large-time asymptotics of these functions at finite temperature. The well-known mapping of the finite-temperature quantum field theory on infinite space to a vacuum theory quantized on the circle leads naturally to correlation functions with imaginary time. A lot is known about these correlation functions, but the analytic continuation to real time is plagued by singularities, so more work is needed in order to investigate the finite-temperature correlation functions in time-like regions. Recent approaches to the problem include a semiclassical method applicable to the regime of small temperature~\cite{SachdevYoungPRL,Sachdev96}, an approach, working in a similar regime, based on identifying the leading singularities of operator matrix elements~\cite{AltshulerKonikTsvelik}, a virial expansion (in powers of soliton density)~\cite{ReyesTsvelik}, and a low-temperature expansion using a regularisation of infinite-volume form factors and an appropriate re-summation scheme \cite{EssKon}. Here, we take a different approach based on solving a famous set of nonlinear partial differential equations (PDEs) satisfied by the thermal correlation functions~\cite{Perk,Lisovyy02,FonsecaZ03}, a generalisation of the nonlinear ordinary differential equations (Painlev\'e III) occurring at zero temperature \cite{McCoy,SMJ,BB,FonsecaZ03}. The main element of these PDEs is the sinh-Gordon PDE which is integrable and possess a zero-curvature formulation, hence is amenable to solution by the classical inverse scattering method (see the books \cite{FadTakBook,BernardBook}). This is essentially a generalization of the Fourier transform to nonlinear PDEs. A linear scattering problem with spectral parameter is associated to the sinh-Gordon equation, such that the associated time-dependent scattering amplitudes (corresponding to the solutions of the scattering problem with incoming plane wave boundary condition at $x\to\infty$ or $x\to-\infty$) are in correspondance with the solutions to the sinh-Gordon equation. The beauty of this approach is that the time evolution of the scattering amplitudes is simple. Hence, finding the scattering amplitudes at zero time (the {\em initial scattering data}) suffices to find them at all times. Then, there are known methods for mapping back these scattering amplitudes to solve the original problem. The difficulties are in calculating the initital scattering data for a given initial sinh-Gordon field configuration, and in mapping the time-evolved scattering amplitudes back to solve the PDEs in a time-like region.

In this work, we first calculate the initial scattering data associated to the thermal equal-time two-point correlation functions in the Majorana theory, by solving a Riemann-Hilbert problem. Our calculation is mainly based on the theory of thermal states and finite-temperature form factors, initiated in \cite{I} in the context of massive integrable QFT (in the context of more general QFT, see \cite{LeplaeUM74,ArimitsuU87,Henning95}). It turns out that the scattering data are essentially given by one-particle finite-temperature form factors of twist fields, evaluated in \cite{I,II}. These scattering data are checked against numerical solutions to the scattering problem and found to be in good agreement. We then recall how these scattering data may be mapped to the two-point correlation functions at all times, via a set of coupled Volterra integral equations known as the Gelfand-Levitan-Marchenko (GLM) equations. The derivation is not entirely standard, because in our case, the solution of the sinh-Gordon equation has complex regions (hence, we really have a hybrid between the sinh-Gordon and sine-Gordon equations). This approach is compared with the known large-distance series expansion for the correlation functions at zero time difference, the form factor expansion in the quantisation on the circle, and shown to reproduce the series term by term for the first few terms. Finally, we solve the GLM equations analytically at leading order (i.e. up to ``exponentially smaller'' terms) in the region $x+t\gg |x-t|, m^{-1},T^{-1}$ (where $x$ is the separation), with $m/T\neq 0$ and $m |x-t|\neq0$, both in the time-like ($t>x$) and space-like ($x>t$) cases. This region is chosen because it illustrates how to ``cross the light cone'' in the GLM equations, and because it hasn't been investigated yet in the literature. We find that at leading order, the two-point function in the disordered regime is given by the analytic continuation of the {\em one-particle} contributions to the form factor expansion in the quantisation on the circle (that is, the one-particle contributions to the finite-temperature form factor expansion), up to a factor of an exponential in $x$ and $t$ that is beyond the reach of the PDE method. The advantage of this approach is that temperature and mass are generic and the set of approximations required to solve the GLM equations analytically are carefully controlled. We believe that the GLM equations we obtained can be used to obtain systematic expansions in any region of space-time, and should give precise numerical solutions.

The paper is organised as follows: in section 2, we present the necessary background including a discussion of the theory of the Majorana fermion and twist fields at finite temperature, the various Hilbert space constructions that we will need, as well as the partial differential equations satisfied by the two point correlation functions of twist fields in the finite-temperature Majorana theory and the associated linear scattering problem. In section 3, we determine the scattering data associated to the linear problem and compare to numerical solutions to the scattering equations. In section 4, we outline the derivation of the GLM equations which allow for the solution to the original problem to be recovered from the time-evolved scattering data. In section 5, we solve the GLM equations for equal-time case and for the region $x+t\gg |x-t|,m^{-1},T^{-1}$. In section 6 we present the results for the correlation functions in the latter region. We conclude in section 7. The appendices contain details of the calculation of the scattering data and further discuss the derivation of the results in section 5.
%%%%%%%%%%%%%%%%%%%%%%%%%%%%%%%%%%%%%%%%%%%%%%%%%%%%%%%%%%
\section{Background}
%%%%%%%%%%%%%%%%%%%%%%%%%%%%%%%%%%%%%%%%%%%%%%%%%%%%%%%%%%
\subsection{The Majorana fermion}\label{SectMajFerm}
%%%%%%%%%%%%%%%%%%%%%%%%%%%%%%%%%%%%%%%%%%%%%%%%%%%%%%%%%%

The theory of massive free Majorana fermions, with real anti-commuting fields $\psi,\b\psi$ on two-dimensional space-time, is defined by the action (we will consider a positive masse $m>0$ throughout)
\beq\label{action}
{\cal A}=i\int{\dd^2 x\left( -\psi(
\partial_x+\partial_t
)\psi+\b{\psi}(
\partial_x-\partial_t
)\overline{\psi}
-m\b{\psi}\psi\right)}\,.
\eeq
Note that this action implies that the leading term of the operator product expansion (OPE) of fermion fields is, as $x\to \pm t$ from the space-like region,
\beq\label{OPEpsipsi}
    \psi(x,t)\psi(0,0) \sim \frc{i}{2\pi(x-t)}~,
    \quad
    \b\psi(x,t)\b\psi(0,0) \sim -
    \frc{i}{2\pi(x+t)}~.
\eeq
We introduce here three quantisation schemes, or rather three Hilbert spaces on which the model can be defined: the Hilbert space of quantisation on the line, that of quantisation on the circle, and the finite-temperature Hilbert space discussed in \cite{I,II}. The latter one is the most important for our purposes.

\subsubsection{Quantisation on the line}

Let's take the theory (\ref{action}) on infinite space $x\in (-\infty,\infty)$, and quantise on equal-time slices $t={\rm constant}$. Then the Hilbert space, ${\cal H}$, is a subspace of the space of field configurations on the line, and is constrained to form a module for the canonical equal-time anti-commutation relations
\beq\label{equaltime}
    \{\psi(x,t),\psi(x',t)\} = \delta(x-x')~,\quad
    \{\b\psi(x,t),\b\psi(x',t)\} = \delta(x-x')~,\quad \{\psi(x,t),\b\psi(x',t)\}=0~.
\eeq
The fermion operators evolve on successive equal-time lines through the equations of motion
\beqa\label{edm}
    \b\p\psi(x,t) &=&
    \frc{m}2 \b\psi(x,t) \n
    \p\b\psi(x,t) &=&
    \frc{m}2 \psi(x,t)\,,
\eeqa
with $\b\p=\frac{1}{2}(\p_x+\p_t)$ and $\p=\frac{1}{2}(\p_x-\p_t)$, and due to the normalisation chosen, they are hermitian, $\psi^\dag = \psi\,, \b\psi^\dag = \b\psi$. A solution is given by:
\beqa
	\psi(x,t) &=& \frc12\sqrt{\frc{m}{\pi}}\,\int \dd\theta\, e^{\theta/2} \lt(A(\theta)\,e^{ip_\theta x - iE_\theta t} + A^\dag(\theta)\,e^{-ip_\theta x + iE_\theta t} \rt) \n
	\b\psi(x,\tau) &=& -\frc{i}2\,\sqrt{\frc{m}\pi}\,\int \dd\theta\, e^{-\theta/2} \lt( A(\theta)\,e^{ip_\theta x - i E_\theta t} -A^\dag(\theta)\,
		e^{-ip_\theta x + i E_\theta t} \rt)~,
\eeqa
where the mode operators $A(\theta)$ and their Hermitian conjugate $A^\dag(\theta)$ satisfy the canonical anti-commutation relations
\beq\label{algtheta}
    \{A^\dag(\theta),A(\theta')\} = \delta(\theta-\theta')~,
\eeq
(other anti-commutators vanishing) and where
\[
    p_\theta = m\sinh\theta~,\quad E_\theta = m\cosh\theta~.
\]
Here $\theta$ is the rapidity, and throughout this work we will also use the variable $\lambda=\exp(\theta)$. The Hilbert space $\cal H$ is simply the Fock space over the algebra (\ref{algtheta}) with vacuum vector $|\vac\ket$ defined by $A(\theta)|\vac\ket=0$. We will use the notation
\beq
    |\theta_1,\ldots,\theta_k\ket = A^\dag(\theta_1)\cdots
    A^\dag(\theta_k)|\vac\ket~.
\eeq
A basis is formed by taking, for instance, $\theta_1>\cdots>\theta_k$. The Hamiltonian is
\beq
	H = \int_{-\infty}^{\infty} \dd\theta\,m\cosh\theta\,A^\dag(\theta) A(\theta)~,
\eeq
and has the property of being bounded from below on $\cal H$ and of generating time translations:
\beq\label{Hpsi}
	i[H,\psi(x,t)] = \frc{\p}{\p t} \psi(x,t)~,\quad i[H,\bar\psi(x,t)] = \frc{\p}{\p t} \bar\psi(x,t)~.
\eeq

\subsubsection{Quantisation on the circle}

Now let's take the theory (\ref{action}) on the circle of circumference $\beta$, $x\in [0,\beta],\;(x=0)\equiv (x=\beta)$. Quantising again on equal-time slices $t={\rm constant}$, the Hilbert space is a subspace of field configurations on the circle. Thanks to the $\Z_2$ symmetry of the action (\ref{action}) under change of signs of the fields, it is possible to define the fermion fields on a double covering of the circle and there are two sectors available for quantisation: periodic or Ramond (R), and anti-periodic or Neveu-Schwarz (NS) conditions around the circle. The Hilbert space is again constrained to form a module for the canonical equal-time anti-commutation relations (\ref{equaltime}), and fermion operators evolving on successive circles obey (\ref{edm}) and are hermitian. We will denote these operators by $\psi_\beta,\,\b\psi_\beta$. With $S={\rm NS}$ and $S={\rm R}$ for the NS and R sectors respectively, and $\Z_{{\rm NS}} = \Z+1/2$ and $\Z_{{\rm R}} = \Z$, the solution is given by
\beqa
    \psi_\beta(x,t) &=& \frc1{\sqrt{2\beta}}
    \,\sum_{n\in\Z_S}\frc{e^{\theta_n/2}}{\sqrt{\cosh\theta_n}}\,
    \lt(
    a_n\,e^{ip_n x - iE_nt} +
    a^\dag_n\,e^{-ip_n x + iE_nt}
    \rt) \n
    \b\psi_\beta(x,t) &=& -\frc{i}{\sqrt{2\beta}}
    \,\sum_{n\in\Z_S}\frc{e^{-\theta_n/2}}{\sqrt{\cosh\theta_n}}\,
    \lt(
    a_n\,e^{ip_n x - iE_nt} -
    a^\dag_n\,e^{-ip_n x + iE_nt}
    \rt)~,
\eeqa
where the discrete mode operators $a_n$ and their Hermitian
conjugate $a^\dag_n$ satisfy the canonical anti-commutation
relations
\beq\label{algn}
    \{a^\dag_n,a_{n'}\} = \delta_{n,n'}~,
\eeq
(other anti-commutators vanishing) and where
\beqa
    p_n &=& m\sinh\theta_n = \frc{2\pi n}\beta ~, \\  E_n &=&
    m\cosh\theta_n~.\no
\eeqa
Here we define the quantised rapidities
\beq\label{thetan}
	\theta_n = {\rm arcsinh}\lt(\frc{2\pi n}{m\beta}\rt)~.
\eeq
The Hilbert space ${\cal H}_\beta^S$ is simply the Fock space over the algebra (\ref{algn}) with vacuum vector $|\vac\ket_\beta^S$ defined by $a_n|\vac\ket_\beta^S=0$. We will use the notation
\beq
    |n_1,\ldots,n_k\ket_\beta^S = a^\dag_{n_1}\cdots
    a^\dag_{n_k}|\vac\ket_\beta^S~.
\eeq
A basis is formed by taking, for instance, $n_1>\cdots>n_k$. The Hamiltonian is given by
\beq\label{Hamilcircle}
    H_\beta^S = {\cal E}_S + \sum_{n\in\Z_S} m\cosh\theta_n\,a^\dag_n a_n
\eeq
and generates time translations:
\beq
	i[H_\beta^S,\psi_\beta(x,t)] = \frc{\p}{\p t} \psi_\beta(x,t)~,\quad i[H_\beta^S,\bar\psi_\beta(x,t)] = \frc{\p}{\p t} \bar\psi_\beta(x,t)~.
\eeq
In (\ref{Hamilcircle}), we have included the vacuum energies ${\cal E}_S$. For both sectors they are infinite, but their difference is finite and given by
\beq
	\Delta{\cal E} \equiv {\cal E}_{{\rm R}} - {\cal E}_{{\rm NS}} =
	\int_{-\infty}^{\infty} \dd\theta \cosh\theta \ln\lt(\frc{1+e^{-m\beta\cosh\theta}}{1-e^{-m\beta\cosh\theta}}\rt)~.
\eeq

\subsubsection{Finite-temperature Hilbert space}

When considering a quantum theory at finite temperature, it is convenient to work with the finite-temperature Hilbert space (or Liouville space) of thermo-field dynamics \cite{LeplaeUM74,ArimitsuU87,Henning95}. For the present model (\ref{action}) on infinite space $x\in(-\infty,\infty)$, this Hilbert space was studied at length in \cite{I,II}. The main idea is to consider (a subspace of) the space ${\rm End}\ ({\cal H})$ of operators on ${\cal H}$, essentially the linear span of all operators formed out of products of any number of operators $a(\theta)$ and $a^\dag(\theta)$ at different rapidities, and to endow it with the inner product (parametrised by the temperature $T$)
\beq\label{ftNS}
	 A\cdot B=\frc{\Tr\lt(e^{-H/T} A^\dag B\rt)}{\Tr\lt(e^{-H/T}\rt)}~,
\eeq
where $A$ and $B$ are any two elements of ${\rm End}\ ({\cal H})$. In fact, like in the case of the quantisation on the circle, thanks to the $\Z_2$ symmetry of the action (\ref{action}), there is another inner product that can be used to provide a Hilbert space structure:
\beq\label{ftR}
	 A\cdot' B=\frc{\Tr\lt(e^{-H/T} \zt A^\dag B\rt)}{\Tr\lt(e^{-H/T} \zt\rt)}~,
\eeq
where $\zt$ is the unitary operator that implements the $\Z_2$ symmetry, $\zt \psi(x,t) \zt = -\psi(x,t)$, etc. (note that it is normalised to $\zt^2=1$). It will be convenient for what follows to have a symbol for the infinite value\footnote{It can be formally evaluated by evaluating the traces explicitly using (zero-temperature) multi-particle states on the line, with a momentum discretisation $\delta p = 2\pi V^{-1}$ where $V$ is the volume, corresponding to $\delta(\theta_1-\theta_2=0)\mapsto mV \cosh\theta_1\;/(2\pi)$}
\beq\label{constZ}
	{\cal Z} = \frc{\Tr\lt(e^{-H/T} \rt)}{\Tr\lt(e^{-H/T} \zt\rt)} = e^{2mV K_1(m/T)/\pi + O(1)}
\eeq
where $V$ is the (infinite) volume. With $S={\rm NS}$ for the Hilbert space $\ft^{{\rm NS}}$ with inner product (\ref{ftNS}) and $S={\rm R}$ for the Hilbert space $\ft^{{\rm R}}$ with inner product (\ref{ftR}), we will use the notation
\[
	{\bf 1}_{{\cal H}} \equiv |\vacft\ket^S,~ \frc{A(\theta)}{g_-^{S}(\theta)}\equiv |\theta\ket_-^S,~ \frc{A^\dag(\theta)}{g_+^{S}(\theta)}\equiv|\theta\ket_+^S,~
%	\frc{a(\theta) a(\theta')}{g_-^{(s)}(\theta) g_-^{(s)}(\theta')}\equiv|\theta, \theta'\ket_{-,-}^s,~
	\frc{A(\theta) A^\dag(\theta')}{g_-^{S}(\theta) g_+^{S}(\theta')}\equiv|\theta, \theta'\ket_{-,+}^S,~\ldots
\]
as well as Dirac's usual bra-ket notation for inner products, where
\beq\label{g}
	g_\pm^{S}(\theta) = \lt\{ \ba{ll} \frc1{1+e^{\mp E_\theta}} & S={\rm NS} \z g_\pm(\theta) \equiv \frc1{1-e^{\mp E_\theta}} & S={\rm R} \ea \rt.
\eeq
(we introduce here the shorter notation $g_\pm(\theta)$ for later convenience). This factor is included here because it provides nicer analytic properties to the matrix elements of twist fields that we will need below. Given an ordering of the rapidities, for instance $\theta_1>\cdots>\theta_k$, all such states form an orthogonal basis, with inner product
\beq
	{\ }^{\hspace{8mm} S}_{\ep_1,\ldots,\ep_k}\bra\theta_1,\ldots,\theta_k|\theta_1',\ldots,\theta_{k'}'\ket^S_{\ep_1',\ldots,\ep_{k'}'} = \delta_{k,k'}
	\prod_{j=1}^k \delta_{\ep_j,\ep_{j'}} \frc{\delta(\theta_j-\theta_j')}{g_{\ep_j}^{S}(\theta_j)}~.
\eeq
Any operator on ${\cal H}$ can also be seen as an operator on $\ft^S$ by, for instance, its left action on operators on $\cal H$. We will denote the left action of any operator $A\in\ft^S$ by $\h{A} \in {\rm End}\ \lt(\ft^S\rt)$. In particular, it is easy to see that
\beq\label{actiona1}
	\h{A}(\theta) |\vacft\ket^S = g_-^{S}(\theta) |\theta\ket_-^S~,\quad \h{A}^\dag(\theta) |\vacft\ket^S = g_+^{S}(\theta) |\theta\ket_+^S
\eeq
and it was shown in \cite{I,II} that, for instance,
\beq\label{actiona2}
	\h{A}(\theta) |\theta'\ket_+^S = g_-^{S}(\theta) |\theta,\theta'\ket_{-,+}^S + \delta(\theta-\theta') |\vacft\ket^S~.
\eeq

%%%%%%%%%%%%%%%%%%%%%%%%%%%%%%%%%%%%%%%%%%%%%%%%%%%%%%%%%%
%%%%%%%%%%%%%%%%%%%%%%%%%%%%%%%%%%%%%%%%%%%%%%%%%%%%%%%%%%
\subsection{Order and disorder fields and their form factors}\label{SecOrdDis}
%%%%%%%%%%%%%%%%%%%%%%%%%%%%%%%%%%%%%%%%%%%%%%%%%%%%%%%%%%
%%%%%%%%%%%%%%%%%%%%%%%%%%%%%%%%%%%%%%%%%%%%%%%%%%%%%%%%%%
As discussed in the introduction, the quantities of interest in the spin chain model (\ref{spinchain}) are correlation functions of the spin matrices themselves. We will be interested in finite-temperature two-point correlation functions $\Tr\lt(e^{-H_I/T} \sigma^{z}_{i}(t)\sigma^{z}_{j}(0)\rt) / \Tr\lt( e^{-H_I/T}\rt)$ in the scaling limit. These correspond to two-point correlation functions of twist fields in the Majorana theory~\cite{KadanoffCeva}; twist fields are the natural local fields associated to the $\Z_2$ symmetry of the theory $\{\psi,\b{\psi}\}\to \{-\psi,-\b{\psi}\}$. This correspondance is a result of writing, in the mapping from the spin model to the Majorana theory, pairs of spin operators, say at sites $i$ and $j$, as exponentials of sums of bilinears in fermionic variables along the segment of the chain joining sites $i$ and $j$. There are two types of twist fields: the order field $\sigma$, and the disorder field $\mu$. Correlation functions of the former give the scaling limit of the spin-chain correlation functions in the ordered regime, whereas those of the latter give correlation functions in the disordered regime.

\subsubsection{Definition}

As operators on the Hilbert space ${\cal H}$ of quantisation on the line, there are two representations of the order twist fields, which we will denote $\sigma_+$ and $\sigma_-$. They may be defined through equal-time exchange relations with the Majorana fermions (see for example~\cite{ItzyksonDrouffe}):
\beqa
    \psi(x)\sigma_+(x') &=& (-1)^{\Theta(x-x')}\sigma_+(x')\psi(x) \quad(x\neq x') \\
    \psi(x)\sigma_-(x') &=& (-1)^{\Theta(x'-x)}\sigma_-(x')\psi(x) \quad(x\neq x')
\eeqa
(and the same holds for $\psi\mapsto\b\psi$), where $\Theta(x)$ is Heavyside's step function. The order fields are further defined by stating that they have a minimal scaling dimension, and that they possess non-zero vacuum expectation values (as operators, they have bosonic statistics). They correspond to the case $g<J$ in the scaling limit of (\ref{spinchain}), for the spin operator we are interested in.

In correlation functions on euclidean space, with insertion of Majorana fermions, this defines fields that create branch cuts on their right ($\sigma_+$) and on their left ($\sigma_-$), when correlation functions are seen as functions of the positions of Majorana fermions. The shape of the cut can be changed without any effect on the correlation functions, as long as it does not cross other local fields that are affected nontrivially by the $\Z_2$ symmetry transformation. At zero temperature, the cut can also be rotated without much effect, so that $\sigma_+$ and $\sigma_-$ are simply related:
\begin{align}
	\bra\vac| \psi(x_1)\cdots &\psi(x_j) \sigma_+(0) \psi(x_1') \cdots \psi(x_k')|\vac\ket\nonumber\\
&=(-1)^k \bra\vac| \psi(x_1)\cdots \psi(x_j) \sigma_-(0) \psi(x_1') \cdots \psi(x_k')|\vac\ket\label{branch}
\end{align}
(and the same equality holds for $\psi\mapsto\b\psi$) -- note that $j+k$ must be even for the correlation function to be nonzero. But such a rotation is more complicated at finite temperature, where in calculating traces, one considers states with both even and odd particle numbers $k$ ($=j$).

The disorder fields $\mu_\pm$ are defined through the leading coefficient of the OPE's $\psi\sigma_\pm$ (as $x\to0$):
\beq\label{OPEps}
	\psi(x)\sigma_\pm(0) \sim \frc{i}{2\sqrt{\pi x + i 0^+}} \mu_\pm(0)~.
\eeq
As fields in euclidean correlation functions, they still have cuts in right/left ($\mu_+/\mu_-$) directions, but as operators on the Hilbert space, they satisfy different exchange relations, because of their fermionic statistics:
\beqa
    \psi(x)\mu_+(x') &=& (-1)^{\Theta(x'-x)}\mu_+(x')\psi(x) \quad(x\neq x') \\
    \psi(x)\mu_-(x') &=& (-1)^{\Theta(x-x')}\mu_-(x')\psi(x) \quad(x\neq x')~.
\eeqa
Their leading OPE coefficients are
\beq\label{OPEpm}
	\psi(x)\mu_\pm(0) \sim \frc{1}{2\sqrt{\pi x + i 0^+}} \sigma_\pm(0)~.
\eeq
The OPEs lead to the following hermiticity properties:
\beq\label{hermit}
	\sigma_\pm^\dag = \sigma_\pm~,\quad \mu_\pm^\dag = \pm\mu_\pm~.
\eeq

Note that operators with branch cut on the right and on the left are simply related to each other through pre- or post-multiplication by the unitary $\Z_2$-symmetry implementing operator $\zt$ introduced in (\ref{ftR}):
\beq\label{ztrel}
	\zt \sigma_\pm = \sigma_\pm \zt = \sigma_\mp ~,\quad \zt\mu_\pm = -\mu_\pm\zt = -\mu_\mp~.
\eeq
Also, we will normalise the operators according to the standard CFT normalisation
\beq\label{CFTnorm}
	\sigma_\pm(x,0)\sigma_\pm(0,0) \sim |x|^{-\frc14}~,\quad
	\mu_\pm(x,0)\mu_\pm(0,0) \sim \pm |x|^{-\frc14}~.
\eeq
Relations (\ref{ztrel}) imply that
\beq\label{normspecial}
	\sigma_+(x,0)\sigma_-(0,0) \sim \zt |x|^{-\frc14}~,\quad
	\mu_+(x,0)\mu_-(0,0) \sim \zt |x|^{-\frc14}
\eeq
(note that these normalisations are different from those used in \cite{I,II}, but more convenient here for clarity).

On a given sector ${\cal H}_\beta^{{\rm NS,R}}$ of quantisation on the circle, operators associated to the fields $\sigma,\mu$ are ill-defined. Rather, these operators are maps from one quantisation sector to another. There are two such maps for every field, ${\cal H}^{{\rm NS}}\to {\cal H}^{{\rm R}}$ and ${\cal H}^{{\rm R}} \to {\cal H}^{{\rm NS}}$, but we will denote both simply by $\sigma$ and $\mu$, as it will be clear from the context which map they represent. Essentially, the cut associated to these twist fields, in this case, should be seen as extending in (imaginary) time, hence changing the quantisation sector at early or at late times.

\subsubsection{Form factors}

The main objects from which asymptotics of the correlation functions we are interested in can be evaluated are matrix elements of twist fields between the vacuum and excited states, or form factors. We recall here the known results concerning these objects (below, $\Or$ is $\sigma$ for even number of particles and $\mu$ for odd number of particles; other form factors are zero).

In the Hilbert space on the line, ${\cal H}$, form factors were evaluated a long time ago \cite{BergKW79} by solving a Riemann-Hilbert problem on the rapidity variables (and then afterwards by a variety of methods), and are given by
\beq
	\frc{\bra \vac | \Or_\pm(x,t) | \theta_1,\ldots,\theta_k\ket}{\bra\vac|\sigma_\pm(0,0)|\vac\ket} = 	(\pm)^k i^{\lt[\frc{k}2\rt]} \lt(\frc{i}{2\pi}\rt)^{\frc{k}2} e^{\sum_{j=1}^k (ixp_{\theta_j} - i t E_{\theta_j})} 
	\prod_{1\leq i<j\leq k} \tanh\lt(\frc{\theta_j-\theta_i}2\rt)~.
\eeq

In the Hilbert space on the circle, ${\cal H}_\beta^{{\rm NS,R}}$, the form factors were evaluated much more recently, first using lattice methods \cite{Bugrij00,Bugrij01}, then using the ``doubling trick'' directly in the QFT model (\ref{action}) \cite{FonsecaZ01}, finally using the connection to finite-temperature form factors in that model \cite{I,II}. Those that will be of interest to us, with the NS vacuum, are given by
\beqa
	\frc{{\ }_\beta^{\hspace{-2mm}{\rm NS}}\bra \vac | \Or(x,t) |n_1,\ldots,n_k\ket_\beta^{{\rm R}}}{
		{\ }_\beta^{\hspace{-2mm}{\rm NS}}\bra\vac|\sigma(0,0)|\vac\ket_\beta^{{\rm R}}}
	&=& i^{\lt[\frc{k}2\rt]} e^{- \Delta{\cal E} it+\sum_{j=1}^k \lt(x\frc{2\pi in_j}{\beta} - i t E_{\theta_{n_j}}\rt)} \times \\ & &
	\prod_{j=1}^k \lt(\sqrt{\frc{2\pi}{\beta E_{\theta_{n_j}}}} h_{+}\lt(\theta_{n_j}+\frc{i\pi}2\rt)\rt) \prod_{1\leq i<j\leq k} \tanh\lt(\frc{\theta_j-\theta_i}2\rt)~, \no
\eeqa
where the function $h_+(\theta)$ is defined below, (\ref{h}).

Finally, on the finite-temperature Hilbert space $\ft^{{\rm NS,R}}$, they were evaluated in \cite{I,II} by deriving a Riemann-Hilbert problem on the rapidity variables, similar to the one for ordinary form factors but with important differences, and by solving it. Those that will be of interest to us, this time on the R Hilbert space, are given by
\beqa\label{ftff}
	\frc{{\ }^{{\rm R}}
		\bra\vacft|\h\Or_\pm(x,t)|\theta_1,\ldots,\theta_k\ket_{\ep_1,\ldots,\ep_k}^{{\rm R}}}{{\ }^{{\rm R}}\bra\vacft|\h\sigma_\pm(0,0)|\vacft\ket^{{\rm R}}}
	&= & (-i)^{k\delta_{\pm,-}} i^{\lt[\frc{k}2\rt]} e^{\mp \Delta{\cal E} x+\sum_{j=1}^k \ep_j(ixp_{\theta_j} - i t E_{\theta_j})} \times \\ & &
	\prod_{j=1}^k h_{\pm\ep_j}(\theta) \prod_{1\leq i<j\leq k} \tanh\lt(\frc{\theta_j-\theta_i}2\rt)^{\ep_i\ep_j} \no
\eeqa
where
\beq\label{h}
	h_\pm(\theta) = \frc{e^{\pm\frc{i\pi}4}}{\sqrt{2\pi}}\,
    \exp\lt[\pm\int_{-\infty\mp i0^+}^{\infty \mp i0^+} \frc{\dd\theta'}{2\pi i}
    \frc1{\sinh(\theta-\theta')}
    \ln\lt(\frc{1+e^{- E_{\theta'}/T}}{1-e^{- E_{\theta'}/T}}\rt)\rt]~.
\eeq
As explained in \cite{I}, finite-temperature form factors specialise both to the zero-temperature ones in the limit $T/m\to0$ (recall that $T$ parametrise the inner product used on $\ft^{{\rm NS,R}}$) when choosing all signes $\ep_j=+$, as is expected, and also to form factors on the circle, through the relations (with time taken imaginary, $t=-i\tau$, for the operator on the finite-temperature Hilbert space):
\beq\label{corres}
	{\ }_\beta^{\hspace{-2mm} {\rm NS}}\bra\vac|\Or(-\tau,-ix)|n_1,\ldots,n_k\ket_\beta^{{\rm R}} = 
	\prod_{j=1}^k \sqrt{\frc{2\pi}{\beta E_{\theta_{n_j}}}}
	{\ }^{{\rm R}}\bra\vacft|\h\Or_+(x,-i\tau)|\theta_{n_1}+\frc{i\pi}2,\ldots,\theta_{n_k}+\frc{i\pi}2\ket_{+,\ldots,+}^{{\rm R}}
\eeq
with $\beta = 1/T$. Hence, finite-temperature form factors offer a way of evaluating form factors on the circle. In a sense, the finite-temperature Hilbert space is the ``analytic continuation'' of the Hilbert space of quantisation on the circle towards positive imaginary eigenvalues of the momentum operator $p\to iE$~\cite{I}. In this sense, the vacua $|\vacft\ket^{{\rm NS,R}}$ should be seen as ``lying at $x\to-\infty$'' and ${\ }^{{\rm NS,R}}\bra\vacft|$ as ``lying at $x\to\infty$''. For instance, the branch cut of $\Or_+(x,i\tau)$ towards the positive $x$ direction on the right-hand side of (\ref{corres}) changes the boundary conditions in that region from periodic (coming from the R finite-temperature Hilbert space) to antiperiodic, as expressed by the NS vacuum of quantisation on the circle on the left-hand side, and it is important that this cut be in the ``direction'' of the vacuum ${\ }^{{\rm R}}\bra\vacft|$ and not of the excited states.

Note that (with all operators at $x=t=0$)
\beq
	{}_\beta^{\hspace{-2mm}{\rm NS}}\bra\vac|\sigma|\vac\ket^{{\rm R}}_\beta = {}_\beta^{{\rm R}}\bra\vac|\sigma|\vac\ket^{{\rm NS}}_\beta
	=  {}^{{\rm R}}\bra\vacft|\h\sigma_\pm|\vacft\ket^{{\rm R}} {\cal Z}^{-\frc12}=
	{}^{{\rm NS}}\bra\vacft|\h\sigma_\pm|\vacft\ket^{{\rm NS}} {\cal Z}^{\frc12} \equiv s_T~,
\eeq
where \cite{Sachdev96,McCoyWu}
\begin{align}
&s_T =\no\\
&m^{\frc18}2^{\frc1{12}} e^{-\frc18} A^{\frc32} \exp\left[ \frc{(m\beta)^2}2
    \int\int_{-\infty}^\infty \frc{\dd\theta_1\dd\theta_2}{(2\pi)^2}
    \frac{\sinh\theta_1\sinh\theta_2}{\sinh(m\beta\cosh\theta_1)\sinh(m\beta\cosh\theta_2)}\times
    \ln\left|\left(\coth\frac{\theta_1-\theta_2}{2}\right)\right|
\right],
\end{align}
and $A$ is Glaisher's constant (and recall that ${\cal Z}$ is given by (\ref{constZ})). The appearance of the constant ${\cal Z}$ is a consequence of the normalisation (\ref{normspecial}) and of the finite-temperature form factor theory developed in \cite{I,II}\footnote{The idea is that it is the two-point functions $ {}^{{\rm R}}\bra\vacft|\h\Or_+ (x,t) \h\Or_-(0,0)|\vacft\ket^{{\rm R}}$ or $ {}^{{\rm NS}}\bra\vacft|\h\Or_+ (x,t) \h\Or_-(0,0)|\vacft\ket^{{\rm NS}}$, with these choices of branch cut directions $\pm$, that one must take to obtain the finite-temperature form factor expansion, hence the clustering on the finite-temperature Hilbert space.}.

%%%%%%%%%%%%%%%%%%%%%%%%%%%%%%%%%%%%%%%%%%%%%%%%%%%%%%%%%%
%%%%%%%%%%%%%%%%%%%%%%%%%%%%%%%%%%%%%%%%%%%%%%%%%%%%%%%%%%
\subsection{Finite temperature correlation functions}\label{sectftcf}
%%%%%%%%%%%%%%%%%%%%%%%%%%%%%%%%%%%%%%%%%%%%%%%%%%%%%%%%%%
%%%%%%%%%%%%%%%%%%%%%%%%%%%%%%%%%%%%%%%%%%%%%%%%%%%%%%%%%%
Whereas zero temperature field theory is often a good approximation in particle physics, condensed matter applications often require the use of finite temperature field theory. This presents a major technical challenge, since  finite temperature correlation functions contain contributions from all states in the Hilbert space. Physical observables at finite temperature are statistical averages over correlation functions, with the Gibbs density matrix $e^{-H/T}$ for a temperature $T$. We therefore consider traces like
\beq
	\braL\cdots \ketL = \frc{\Tr\lt( e^{-H/T} \cdots\rt)}{\Tr\lt(e^{-H/T}\rt)}~.\label{EqTrace}
\eeq
In order to calculate such quantities, one may use two Hilbert spaces: that of quantisation on the circle, and the finite-temperature Hilbert space.

The first arises form the well known ``mapping to the cylinder''. With the euclidean signature (i.e. in imaginary time), finite-temperature correlation functions of quantum field theory models in one infinite dimension are equivalent to vacuum correlations of the associated model quantised on the circle of diameter $\beta=1/T$. That is, finite temperature correlation functions may be written as
\beq
    \braL \Or(x,-i\tau) \cdots \ketL = {\cal N}_{S,S'} \big(e^{i\pi s/2}\cdots\big)\,{\ }_\beta^{S}\bra\vac| \Or_\beta(-\tau,-ix)
    \cdots |\vac\ket_\beta^{S'}~, \label{EqCircle}
\eeq
where $s$ is the spin of $\Or$ and ${\cal N}_{S,S'}$ is some normalisation constant. The operator $\Or_\beta(-\tau,-ix)$ is the corresponding operator in quantisation on the circle, with space variable $-\tau$ (parameterizing the circle of circumference $\beta$) and Euclidean time variable $x$ (on the line). The sectors $S,S'$ depend on the operators $\Or,\ldots$ inserted. The trace (\ref{EqTrace}) with insertion of operators that are local with respect to the fermion fields naturally corresponds to the NS sector, $S=S'={\rm NS}$, due to the Kubo-Martin-Schwinger (KMS) identity
\begin{gather}\label{KMS}
    \braL \Or(x,t) \cdots \ketL =
    (-1)^{f_\Or} \braL \Or(x,t-i\beta) \cdots \ketL ~,
\end{gather}
where $f_\Or$ is 1 for an operator with fermionic statistics, and 0 otherwise. In this case, ${\cal N}_{{\rm NS},{\rm NS}} = 1$. This is the sector with the lowest vacuum energy. However, with insertion of twist fields, if the branches are chosen so that at $x\to+\infty$ there is a branch cut, then we will have $S={\rm R}$, and if they are chosen so that at $x\to-\infty$ there is a branch cut, then we will have $S'={\rm R}$, with a factor ${\cal Z}^{-\frc12}$ in the number ${\cal N}_{S,S'}$ for every branch cut. In particular, if the unitary operator $\zt$ is inserted, other insertions being local with respect to the fermion fields, then $S=S'={\rm R}$, with ${\cal N}_{{\rm R},{\rm R}} = {\cal Z}^{-1}$ and $\zt_\beta = 1$. From the viewpoint of the model (\ref{spinchain}), correlation functions correspond to choosing the branch so that the branch cut only extends between the twist fields, so that $S=S'={\rm NS}$.

The representation (\ref{EqCircle}) leads to a large-distance expansion of finite-temperature correlation functions, through insertion of the resolution of the identity on the Hilbert space ${\cal H}_\beta^{{\rm NS}}$. We write it here for twist fields $\Or=\sigma$ or $\Or=\mu$ in the quantisation on the circle in the NS sector:
\begin{gather}
    {}_\beta^{\hspace{-2mm}{\rm NS}}\bra\vac|\Or(-\tau,-ix)\Or(0,0)|\vac\ket_\beta^{{\rm NS}} =  \sum_{k=0}^\infty
    \sum_{n_1,\ldots,n_k\in\Z} \frc{e^{-\Delta{\cal E}x + \sum_{j=1}^k \lt(-\tau \frc{2\pi i n_j }{\beta} - x E_{\theta_{n_j}}\rt)}}{k!} \label{circleffexp}\\
    \phantom{{}_\beta^{\hspace{-2mm}{\rm NS}}\bra\vac|\h\Or(-\tau,-ix)\Or(0,0)|\vac\ket_\beta = }{}\times
    {\ }_\beta^{\hspace{-2mm}{\rm NS}}\bra\vac|\Or(0,0)|n_1,\ldots,n_k\ket_\beta^{{\rm R}}
    {\ }_\beta^{{\rm R}}\bra n_1,\ldots,n_k|\Or(0,0)|\vac\ket_\beta^{{\rm NS}}~.\nonumber
\end{gather}

The second way of dealing with finite-temperature correlation functions is using the finite-temperature Hilbert space. From the basic definition of this Hilbert space, it is easy to see that
\beq
    \braL \Or(x,t) \cdots \ketL = {\ }^{{\rm NS}}\bra\vacft| \h\Or(x,t)
    \cdots |\vacft\ket^{{\rm NS}}~. \label{EqFiniteT}
\eeq
Although the NS sector naturally appears here, it is possible, when twist fields are inserted, to use the R sector as well. Since we are interested in configurations giving a branch cut that extends only between the two twist fields $\Or=\sigma$ or $\Or=\mu$, we may use
\beq\label{NSR}
	{\ }^{{\rm NS}}\bra\vacft| \h\Or_+(x,t) \h\Or_+(0,0) |\vacft\ket^{{\rm NS}} =
	{\cal Z}^{-1} {\ }^{{\rm R}}\bra\vacft| \h\Or_+(x,t) \h\Or_-(0,0) |\vacft\ket^{{\rm R}}~.
\eeq
This can be deduced from the relations (\ref{ztrel}). One can then use a resolution of the identity on $\ft^{{\rm R}}$ in order to obtain a large-distance expansion of the finite-temperature correlation functions, for $x>0$:
\beqa\label{finiteTffexp} &&
	{\ }^{{\rm R}}\bra\vacft| \h\Or_+(x,t) \h\Or_-(0,0) |\vacft\ket^{{\rm R}} = \\ && \qquad
        \;\sum_{k=0}^\infty \sum_{\ep_1,\ldots,\ep_k\in\{\pm\}}
    \int_{\{{\rm Im}(\theta_j)=\ep_j 0^+\}} \frc{\dd\theta_1\cdots
    \dd\theta_k\;
    e^{-\Delta{\cal E} x + \sum_{j=1}^k\ep_j (ix p_{\theta_j}-i t E_{\theta_j})}}{
    k!\;\prod_{j=1}^k\lt(1-e^{-\ep_j
    E_{\theta_j}/T}\rt)}\;\times \n && \qquad\qquad\qquad \times\;
	{\ }^{{\rm R}}\bra\vacft|\h\Or_+(0,0)|\theta_1,\ldots,\theta_{k}\ket^{{\rm R}}_{\ep_1,\ldots,\ep_{k}}
	{\ }^{\hspace{8mm} {\rm R}}_{\ep_1,\ldots,\ep_k}\bra\theta_1,\ldots,\theta_k|\h\Or_-(0,0)|\vacft\ket^{{\rm R}}~.\no
\eeqa
It is important, in deriving this expansion, that the cuts of the twist fields extend towards the ``direction'' of the vacuum states (see the discussion after (\ref{corres})); this is why we need to use the Ramond Hilbert space. It is also important that $x>0$, although the region $x<0$ can be obtained easily by complex conjugation. By deformation of contours, this expansion along with (\ref{NSR}) can be seen to reproduce (\ref{circleffexp}) \cite{AltshulerKonikTsvelik,II}.

Both expansions above are convergent whenever $x^2>t^2$ (and $x>0$), with $t={\rm Im}(\tau)$ in (\ref{circleffexp}). The expansion (\ref{finiteTffexp}) may offer the hope of a series converging in time-like regions by deformation of contours, but this process is plagued by singularities, except for the one-particle contributions. In the following sections, we will show that the combination of the finite-temperature Hilbert space approach with the integrable nonlinear PDEs satisfied by twist-field correlation functions allows one to investigate time-like regions.

%%%%%%%%%%%%%%%%%%%%%%%%%%%%%%%%%%%%%%%%%%%%%%%%%%%%%%%%%%
%%%%%%%%%%%%%%%%%%%%%%%%%%%%%%%%%%%%%%%%%%%%%%%%%%%%%%%%%%
\subsection{Correlation functions and integrable differential equation}
%%%%%%%%%%%%%%%%%%%%%%%%%%%%%%%%%%%%%%%%%%%%%%%%%%%%%%%%%%
%%%%%%%%%%%%%%%%%%%%%%%%%%%%%%%%%%%%%%%%%%%%%%%%%%%%%%%%%%
The scaling limit of finite-temperature two-point correlation functions of spin matrices in the quantum Ising chain gives, depending on the regime, two finite-temperature correlation functions in the Majorana theory:
\begin{align}
	G(x,t) = \bra \sigma_+(x,t) \sigma_+(0,0) \ket_T & \qquad \mbox{(ordered regime)} \\
	\t{G}(x,t) = \bra \mu_+(x,t) \mu_+(0,0)\ket_T & \qquad \mbox{(disordered rgime).}
\end{align}
At zero temperature, these correlation functions satisfy a well known set of ordinary differential equations~\cite{McCoy,SMJ,BB,FonsecaZ03}. These equations can be generalised to partial differential equations at finite temperature \cite{Perk,Lisovyy02,FonsecaZ03}. The resulting PDEs satisfied by the correlation functions $G$ and $\tilde{G}$ can be expressed in terms of $\varphi$ and $\chi$, defined as
\begin{align}
&G \equiv s_T^2 e^{\chi/2} \cosh(\varphi/2)\nonumber\\
&\t{G} \equiv s_T^2 e^{\chi/2} \sinh(\varphi/2)~.\label{EqGGtilde}
\end{align}
In terms of these functions, the PDEs are
\begin{align}
&\p\b\p \varphi = \frc{m^2}8 \sinh(2\varphi)\label{shG}\\
&\p\b\p \chi = \frc{m^2}8 (1-\cosh(2\varphi))\label{eqChi}\\
\p^2\chi + (\p\varphi)^2 &= \b\p^2\chi + (\b\p\varphi)^2 = 0~,\label{eqOthers}
\end{align}
which include the sinh-Gordon equation for $\varphi$.

The sinh-Gordon equation (\ref{shG}) may be rewritten as the following linear problem with spectral parameter $\lambda=e^{\theta}$ \cite{FadTakBook,BernardBook}
\beq\label{linprob}
	(\p_x - A_x) \Psi = (\p_t - A_t) \Psi = 0~,
\eeq
where $\Psi(x,t)$ is a well-defined 2-component vector-valued function of $x$ and $t$, and where
\beqa
	A_x &=& \frc{i}4 \mato{cc} 2i \p_t\varphi & m (\lambda e^{-\varphi} - \lambda^{-1} e^{\varphi}) \\
		m (\lambda e^{\varphi} - \lambda^{-1} e^{-\varphi}) & -2i \p_t \varphi \matf \\
	A_t &=& \frc{i}4 \mato{cc} 2i \p_x\varphi & -m (\lambda e^{-\varphi} + \lambda^{-1} e^{\varphi}) \\
		-m (\lambda e^{\varphi} + \lambda^{-1} e^{-\varphi}) & -2i \p_x \varphi \matf~.
\eeqa
Well-definiteness of $\Psi(x,t)$ for any initial conditions (and for any spectral parameter) implies that $\p_t A_x - \p_x A_t = [A_t, A_x]$, which implies (\ref{shG}).

In fact, under evolution in real time, the function $\varphi$ is expected to acquire complex values in time-like regions of space-time. This is a simple consequence of the fact that quantum operators do not commute when separated by time-like distances in relativistic quantum field theory. Hence, we really have a ``hybrid'' between the sinh-Gordon equation and the sine-Gordon equation. From the viewpoint of the PDE above, the complex nature of $\varphi$ for a real initial condition at $t=0$ can be seen as a consequence of the singularity at $x=0$ of this initial condition. This singularity comes from the fact that $\lim_{x\to0} \t{G}(x,0)/G(x,0) = 1$, and stays on the light cone afer time evolution. In order to properly define the solution to the PDE, we need a prescription for going through it. The natural prescription, to which our results will apply, is Wick's rotation (Feynman's prescription) ${\rm Im}(t) = 0^-$. This corresponds to the usual definition of real-time correlation functions as analytic continuations from those in imaginary time $t=-i\tau,\;\tau\in\R$, where they stay real for all real $\tau$ (that is, where operators are always at space-like distances).

Hence at imaginary times $t=-i\tau$ (with $\p_t \varphi = i\p_\tau\varphi$), the function $\varphi$ is real. We will make use of two symmetries enjoyed by the linear equation $(\p_x - A_x(x;\theta)) \Psi(x;\theta) = 0$ (recall that we use $\lambda=e^{\theta}$), leading to two transforms in the space of solutions, $\Psi\mapsto \b\Psi$ and $\Psi\mapsto \Psi^\shift$, with:
\beqa
	A_x^*(x;\theta) = -A_x(x) &\Rightarrow& \b\Psi(x;\theta) = \Psi^*(-x;\theta) \n
	A_x(x;\theta+i\pi) = \sigma_3 A_x(x,\theta) \sigma_3 &\Rightarrow& \Psi^\shift(x;\theta) = \sigma_3\Psi(x;\theta+i\pi)~.\no
\eeqa
Note that only the second is a symmetry in real time.

%%%%%%%%%%%%%%%%%%%%%%%%%%%%%%%%%%%%%%%%%%%%%%%%%%%%%%%%%%
%%%%%%%%%%%%%%%%%%%%%%%%%%%%%%%%%%%%%%%%%%%%%%%%%%%%%%%%%%
\section{Determining the scattering data}
%%%%%%%%%%%%%%%%%%%%%%%%%%%%%%%%%%%%%%%%%%%%%%%%%%%%%%%%%%
%%%%%%%%%%%%%%%%%%%%%%%%%%%%%%%%%%%%%%%%%%%%%%%%%%%%%%%%%%
The classical inverse scattering method \cite{FadTakBook,BernardBook} is a generalisation of the Fourier transform to the case of nonlinear partial differential equations with a zero curvature formulation. The sinh-Gordon PDE above has such a formulation, the linear problem, and hence is amenable to solution by this method. Let us consider solutions to the spatial part of the linear problem with incoming plane wave asymptotic conditions from $x=\infty$ (positive frequency), or from $x=-\infty$ (negative frequency). It can be seen that these exist thanks to the fact that the function $\varphi$ vanishes exponentially at $x\to\pm\infty$. The two corresponding solutions are known as the Jost solutions, which we will denote by $\Psi_{\pm}$, and form a linearly independent set of solutions that are analytic in ${\rm Im}(\theta) \in [0,\pi]$, which corresponds to the upper half plane of $\lambda$. Let us focus on one of them:
\beqa
	\Psi_{+} &\stackrel{x\to\infty}\sim& e^{\frc{ip_\theta x}2} \mato{c} 1\\1 \matf \n
		&\stackrel{x\to-\infty}\sim& a e^{\frc{ip_\theta x}2} \mato{c}1\\1\matf - be^{-\frc{ip_\theta x}2} \mato{c} 1\\-1 \matf~.
\eeqa
The functions $a(\theta)$ (called the Jost function) and $b(\theta)$ are known as the scattering data. Given any $\varphi$ as function of $x$ (with appropriately vanishing large-$|x|$ behaviours), the linear problem above fixes the scattering data. The classical inverse scattering method gives the inverse map: given the scattering data, a function $\varphi$ of $x$ can be determined. This is done by solving set of integral equations known as the Gelfand-Levitan-Marchenko (GLM) equations. Through the time part of the linear problem (\ref{linprob}), the scattering data acquire a simple time dependence:
\beq\label{timeevol}
	a(\theta , t)=a(\theta) ~, \quad b(\theta ,t)=e^{iE_\theta t} b(\theta)~.
\eeq
Their time evolution gives, by the inverse scattering, functions $\varphi(x,t)$ that solve the sinh-Gordon equation (\ref{shG}). The Jost function $a(\theta)$ does not have poles in ${\rm Im}(\theta) \in [0,\pi]$, since it is proportional to the Wronskian of the Jost solutions, $a(\theta)=-{\rm det}(\Psi_+,\Psi_-)/2$. However, $b(\theta)$ may have poles in that strip. The presence of these poles is related to the way in which the function $\varphi$ vanishes as $|x|\to\infty$. 

The solution to (\ref{shG}) that corresponds to the correlation function should be uniquely fixed by requiring that $\phi(x,t-iT^{-1}) = \phi(x,t)$ and by requiring appropriate asymptotic conditions for $x\to\infty$ and $x^2\to t^2$. These conditions are at the basis of the Riemann-Hilbert problem that fixes the finite-temperature form factors evaluated in \cite{I,II}. In this section, we will determine the initial scattering data that corresponds to the correlation function by direcly using the finite-temperature form factors. Then, in section~\ref{SecGLMOv}, we will explain how the correlation functions are obtained from the GLM equations.

%%%%%%%%%%%%%%%%%%%%%%%%%%%%%%%%%%%%%%%%%%%%%%%%%%%%%%%%%%
%%%%%%%%%%%%%%%%%%%%%%%%%%%%%%%%%%%%%%%%%%%%%%%%%%%%%%%%%%
\subsection{The special solution}\label{SecSpSol}
%%%%%%%%%%%%%%%%%%%%%%%%%%%%%%%%%%%%%%%%%%%%%%%%%%%%%%%%%%
%%%%%%%%%%%%%%%%%%%%%%%%%%%%%%%%%%%%%%%%%%%%%%%%%%%%%%%%%%
In order to obtain the scattering data, a pair of independent solutions to the spatial part of the linear problem are required (or at least both their $x\to\infty$ and $x\to-\infty$ asymptotics). In this sub-section, we show how to evaluate both asymptotics at large $|x|$ of one solution which is related to objects in the QFT model.

The objects in the QFT model leading to a particular solution to the linear problem (\ref{linprob}) are related to the following traces (recall the notation (\ref{EqTrace})):
\beqa
	F(x,t;\theta) &\equiv& \lt\braL \sigma_+\lt(\frc{x}2,\frc{t}2\rt) A^\dag(\theta) \;\mu_+\lt(-\frc{x}2,-\frc{t}2\rt) \rt\ketL \\
	\t{F}(x,t;\theta) &\equiv& \lt\braL \mu_+\lt(\frc{x}2,\frc{t}2\rt) A^\dag(\theta) \;\sigma_+\lt(-\frc{x}2,-\frc{t}2\rt) \rt\ketL~.
\eeqa
It can be shown that
\beq
	\Psi = \Psi_{\sym} \equiv e^{-\chi/2} \bra\sigma\ket_T^{-2} \mato{c} \t{F} -iF \\ \t{F} +iF \matf
\eeq
is a solution to (\ref{linprob}), with $\lambda = e^{\theta}$. The calculation is included in appendix~\ref{SecDerivSp}, and follows closely that of \cite{FonsecaZ03}.

In order to solve its associated connection problem (i.e. the problem of finding the asymptotic form at $x\to-\infty$ knowing that at $x\to\infty$, or vice versa), we can evaluate the large-$|x|$ asymptotics using finite-temperature form factors. Since only the initial scattering data are necessary, let us consider the case $t=0$. Using a resolution of the identity on the Hilbert space $\ft^{{\rm R}}$, just as in equation~(\ref{finiteTffexp}), we have, as $x\to\infty$,
\beqa
	F(x,0;\theta) &=& {\cal Z}^{-1} {\ }^{{\rm R}}\bra\vacft| \h\sigma_+\lt(\frc{x}2,0\rt) \h{A}^\dag(\theta) \; \h\mu_-\lt(-\frc{x}2,0\rt) |\vacft\ket^{{\rm R}} \n
	 &\sim& {\cal Z}^{-1} {\ }^{{\rm R}}\bra\vacft| \h\sigma_+\lt(\frc{x}2,0\rt) |\vacft\ket^{{\rm R}}
		{\ }^{{\rm R}}\bra\vacft| \h{A}^\dag(\theta) \;\h\mu_-\lt(-\frc{x}2,0\rt) |\vacft\ket^{{\rm R}} \n
	&=& {\cal Z}^{-1} {\ }^{{\rm R}}\bra\vacft| \h\sigma_+\lt(\frc{x}2,0\rt) |\vacft\ket^{{\rm R}}
		\;g_-(\theta) {\ }_-^{{\rm R}}\bra \theta| \h\mu_-\lt(-\frc{x}2,0\rt) |\vacft\ket^{{\rm R}} \n
	&=& -i s_T^2 e^{-\Delta {\cal E} x-\frc{ip_\theta x}2}\,g_-(\theta) h_-(\theta) ~.
\eeqa
Here, we use the functions (\ref{g}) and (\ref{h}). In the first step, we inserted $\zt^2=1$ in $\braL\cdots\ketL$ and used (\ref{ztrel}) in order to have the Ramond Hilbert space $\ft^{{\rm R}}$, in the second step we inserted the first term of the resolution of the identity, which gives the leading large-$x$ asymptotic, in the third step we used equation (\ref{actiona1}) and in the last step we used the complex conjugate of the form factor formula (\ref{ftff}) for $k=1$ and for the minus sign (with $h_\pm(\theta)^* = h_\mp(\theta)$), along with the anti-hermitian property (\ref{hermit}) for $\mu_-$. It can be checked that inserting the resolution of the identity between the operators $\h{A}^\dag$ and $\mu_-$, using (\ref{actiona2}), gives the same result.

The asymptotic form of $\t{F}$ follows similarly, for $x\to\infty$:
\beqa
	\t{F}(x,0;\theta) &=& {\cal Z}^{-1} {\ }^{{\rm R}}\bra\vacft| \h\mu_+\lt(\frc{x}2,0\rt) \h{A}^\dag(\theta) \;\h\sigma_-\lt(-\frc{x}2,0\rt) |\vacft\ket^{{\rm R}} \n
	&\sim& {\cal Z}^{-1} {\ }^{{\rm R}}\bra\vacft| \h\mu_+\lt(\frc{x}2,0\rt) \h{A}^\dag(\theta) |\vacft\ket^{{\rm R}}
		{\ }^{{\rm R}}\bra\vacft| \h\sigma_-\lt(-\frc{x}2,0\rt) |\vacft\ket^{{\rm R}} \n
	&=& s_T^2 e^{-\Delta {\cal E} x+\frc{ip_\theta x}2} \,g_+(\theta) h_+(\theta)~.
\eeqa
For the asymptotics as $x\to-\infty$, we take complex conjugates in order to have operators ordered form left to right with decreasing values of their positions (so that the finite-temperature form factor expansion can be used),
\beq
	F(x,0;\theta)^* =   \lt\braL \mu_+\lt(-\frc{x}2,0\rt) A(\theta) \sigma_+\lt(\frc{x}2,0\rt) \rt\ketL ~.
\eeq
Now this expression can be evaluated as $x\to-\infty$, as above,
\beq
	F(x,0;\theta)^* \sim s_T^2 e^{\Delta {\cal E} x+\frc{ip_\theta x}2} \,g_-(\theta) h_-(\theta)
\eeq
which gives
\beq
	F(x,0;\theta) \sim s_T^2 e^{\Delta {\cal E} x-\frc{ip_\theta x}2} \,g_-(\theta) h_+(\theta)~.
\eeq
Similarly, for $x\to-\infty$,
\beq
	\t{F}(x,0;\theta) \sim is_T^2 e^{\Delta {\cal E} x+\frc{ip_\theta x}2} \,g_+(\theta) h_-(\theta)~.
\eeq

In summary,  putting back the (imaginary) time dependence, the asymptotic forms as $|x|\to\infty$ of the solution coming from QFT is
\beqa
	\Psi_\sym &\stackrel{x\to\infty}=& e^{\frc{ip_\theta x}2- \frc{E_\theta \tau}2} g_+ h_+ \mato{c} 1\\1 \matf
		- e^{-\frc{ip_\theta x}2+ \frc{E_\theta\tau}2} g_- h_- \mato{c} 1\\-1 \matf \\
	&\stackrel{x\to-\infty}=& ie^{\frc{ip_\theta x}2 - \frc{E_\theta\tau}2} g_+ h_- \mato{c} 1\\1 \matf
		- ie^{-\frc{ip_\theta x}2+ \frc{E_\theta\tau}2} g_- h_+ \mato{c} 1\\-1 \matf ~,
\eeqa
where the real exponential factor $e^{\pm \Delta{\cal E}x}$ was absorbed into the function $\chi$. Note that this solution is not a Jost solution, but is the most symmetric one:
\beq
	\b\Psi_\sym = -i\Psi_\sym ~,\quad \Psi_\sym^\shift = -i \Psi_\sym^\shift~.
\eeq
%%%%%%%%%%%%%%%%%%%%%%%%%%%%%%%%%%%%%%%%%%%%%%%%%%%%%%%%%%
%%%%%%%%%%%%%%%%%%%%%%%%%%%%%%%%%%%%%%%%%%%%%%%%%%%%%%%%%%
\subsection{The other solution and the scattering data}\label{SecScattDat}
\begin{figure}[htbp]
  \vspace{9pt}
  \centerline{\hbox{ \hspace{0.0in}
    \epsfxsize=3in
    \epsffile{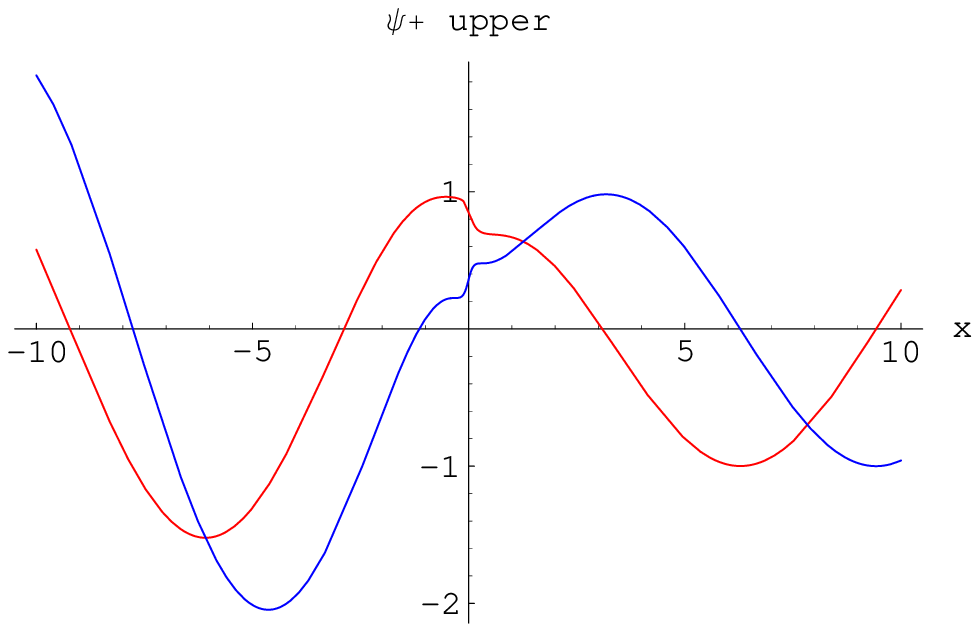}
    \hspace{0.25in}
    \epsfxsize=3in
    \epsffile{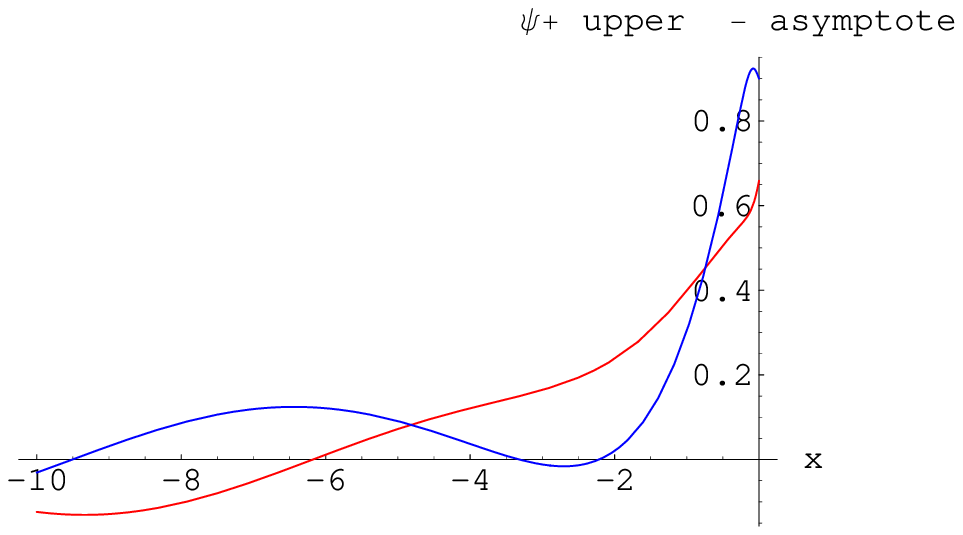}
    }
  }
  \vspace{4pt}
  \hbox{\hspace{1.3in} (a) \hspace{1.70in} (b)}
  \vspace{9pt}
  \caption{Comparison of the scattering data to numerical solution. The numerical solution was obtained by first evaluating the function $\varphi$ using its form factor expansion in the quantisation on the circle, (\ref{circleffexp}), up to 6-particle contributions, then by solving numerically (\ref{linprob}) from incoming plane wave behaviour at $x=10$. (a) shows the real (red) and imaginary (blue) parts of the first (upper) component of the Jost solution $\Psi_+$. This should be compared to (b), showing the difference from the scattering data prediction, which is supposed to be exact for $x\rightarrow -\infty$. The simulation is for $p_\theta =1$ and $m=T=1$. The results are in good agreement (to 5\% for $x=-10$).}
  \label{figNumData}
\end{figure}

%%%%%%%%%%%%%%%%%%%%%%%%%%%%%%%%%%%%%%%%%%%%%%%%%%%%%%%%%%
%%%%%%%%%%%%%%%%%%%%%%%%%%%%%%%%%%%%%%%%%%%%%%%%%%%%%%%%%%
In order to calculate the initial scattering data corresponding to the correlation functions, the asymptotics of a second, linearly independent solution to the linear problem is needed. In this sub-section, we find these asymptotics and hence derive the scattering data. Inspired by the form of the symmetric solution of the previous section, let us look for a complete set of solutions with asymptotics (again, at $\tau=0$)
\beqa
	\Psi_1 &\stackrel{x\to\infty}=& \alpha e^{\frc{ip_\theta x}2} g_+ h_+ \mato{c} 1\\1 \matf
		- \beta e^{-\frc{ip_\theta x}2} g_- h_- \mato{c} 1\\-1 \matf \\
	&\stackrel{x\to-\infty}=& ie^{\frc{ip_\theta x}2 } g_+ h_- \mato{c} 1\\1 \matf \\
	\Psi_2 &\stackrel{x\to\infty}=& (1-\alpha) e^{\frc{ip_\theta x}2} g_+ h_+ \mato{c} 1\\1 \matf
		- (1-\beta) e^{-\frc{ip_\theta x}2} g_- h_- \mato{c} 1\\-1 \matf \\
	&\stackrel{x\to-\infty}=& - ie^{-\frc{ip_\theta x}2} g_- h_+ \mato{c} 1\\-1 \matf ~.
\eeqa
The quantities $\alpha$ and $\beta$ are functions of the spectral parameter $\theta$ to be determined, and the asymptotics as $x\to\infty$ of the two solutions are chosen to be in agreement with the symmetric solution of the previous section. The Jost solutions may then be obtained by a linear combination of the equations above, and the scattering data obtained:
\beq\label{scdata1}
	a = \frc{i (\beta-1) h_-}{(\beta-\alpha)h_+}~,\quad b = \frc{i\beta g_-}{(\beta-\alpha)g_+}~.
\eeq
Since $\Tr A_x=0$, the Wronskian ${\rm det}(\Psi_1, \Psi_2)$ is independent of $x$. Using this, we find one constraint:
\beq\label{eqab0}
	\beta = 1+\alpha~.
\eeq
The symmetry-transformed solutions, $\b{\Psi}$ and $\Psi^\shift$, also satisfy the linear problem. Hence,
\beqa
	\b\Psi_1 &\stackrel{x\to\infty}=& -i e^{\frc{ip_\theta x}2} g_+ h_+ \mato{c}1\\1\matf \n
		&\stackrel{x\to-\infty}=& \alpha^* e^{\frc{ip_\theta x}2} g_+ h_- \mato{c}1\\1\matf - \beta^*e^{-\frc{ip_\theta x}2} g_- h_+ \mato{c} 1\\-1 \matf\no
\eeqa
and by the fact that $\Psi_1$ and $\Psi_2$ form a complete set of solutions, we find, from looking at the $x\to-\infty$ asymptotics,
\beq
	\b\Psi_1 = -i\alpha^* \Psi_1 -i\beta^* \Psi_2~.
\eeq
This gives, from the $x\to\infty$ asymptotics,
\beq\label{eqab1}
	1=\alpha^*\alpha + \beta^*(1-\alpha) ~,\quad 0=\alpha^*\beta + \beta^*(1-\beta)~.
\eeq
Also, we have
\beqa
	\Psi^\shift_1 &\stackrel{x\to\infty}=& -i\beta(\theta+i\pi) e^{\frc{ip_\theta x}2} g_+ h_+ \mato{c} 1\\1 \matf
			+i\alpha(\theta+i\pi) e^{-\frc{ip_\theta x}2} g_- h_- \mato{c} 1\\-1 \matf \n
		&\stackrel{x\to-\infty}=& - e^{-\frc{ip_\theta x}2} g_- h_+ \mato{c} 1\\-1 \matf ~.\no
\eeqa
That is, $\Psi^\shift_1 = -i\Psi_2$, from which we have
\beq\label{eqab2}
	1-\alpha(\theta) =\beta(\theta+i\pi)~,\quad 1-\beta(\theta) = \alpha(\theta+i\pi)~.
\eeq
Equations (\ref{eqab0}) and (\ref{eqab1}) show that $\alpha \in \R$ (for $\theta\in\R$), and (\ref{eqab0}) with (\ref{eqab2}) gives
\beq\label{eqalpha}
	\alpha(\theta+i\pi) = -\alpha(\theta)~.
\eeq
A large-$\theta$ analysis (see the appendix~\ref{SecAsym}) leads to the condition
\beq
	\alpha(\theta) \sim 1 \quad {\rm as}\quad \theta\to\pm\infty~.
\eeq
The final constraint comes from the fact that $a(\theta)$ is analytic in the strip ${\rm Im}(\theta)\in [0,\pi]$ -- recall from above that this follows from the analyticity of the solutions $\Psi_\pm$ in this strip and from the Wronskian identity $a(\lambda)=-{\rm det}(\Psi_+,\Psi_-)/2$. Using the analytic structure of the function $h_-(\theta)$ in the strip ${\rm Im}(\theta)\in[-\pi,\pi]$ (with $\theta_n$ defined in (\ref{thetan}) where $\beta=1/T$)
\beqa
	h_- &:& \mbox{poles}\quad \theta = \theta_n + \frc{i\pi}2~,\quad n \in \Z + \frc12 \n
		&& \mbox{zeroes} \quad \theta = \theta_n + \frc{i\pi}2~,\quad n \in \Z ~,\no
\eeqa
the conditions above form a Riemann Hilbert problem for the function $\alpha(\theta)$ which can be stated as follows:
\begin{itemize}
\item $\alpha(\theta)\in\R$ for $\theta\in\R$
\item $\alpha(\theta+i\pi) = -\alpha(\theta)$
\item $\alpha(\theta) \sim 1$ as $\theta\to\pm\infty$
\item $\alpha(\theta)$ has zeroes at $\theta = \frc{i\pi}2+\theta_n$ for $n\in\Z+\frc12$
\item $\alpha(\theta)$ is analytic for ${\rm Im}(\theta) \in [0,\pi]$ except possibly for poles at $\theta = \frc{i\pi}2+\theta_n$ for $n\in\Z$
\end{itemize}
The unique solution is:
\beq
	\alpha(\theta) = \frc{1+e^{-E_\theta/T}}{1-e^{-E_\theta/T}}\quad \Rightarrow \quad \beta(\theta) = \frc{2}{1-e^{-E_\theta/T}} = 2g_+(\theta)~,
\eeq
and the resulting analytic structure of the function $\alpha(\theta)$ in the strip ${\rm Im}(\theta)\in[-\pi,\pi]$ is
\beqa
	\alpha &:& \mbox{poles} \quad \theta = \theta_n \pm \frc{i\pi}2~,\quad n \in \Z \n
		&& \mbox{zeroes} \quad \theta = \theta_n \pm \frc{i\pi}2~,\quad n \in \Z +\frc12~. \no
\eeqa
These results have been shown to be in good agreement with numerical solutions to the linear differential equations, see figure~\ref{figNumData}.

Then, the scattering data (\ref{scdata1}) can be simplified further, using the identities $h_-(\theta)h_+(\theta) = (2\pi\alpha(\theta))^{-1}$ (the functions $h_\pm(\theta)$ are related by complex conjugation, and also by $h_+(\theta+i\pi) = ih_-(\theta)$), to
\beq\label{scdata}
	a(\theta) = \frc{i}{2\pi h_+(\theta)^2}~,\quad b(\theta) = 2ig_-(\theta)
\eeq
(recall the functions $g_\pm(\theta)$ and $h_\pm(\theta)$,  (\ref{g}) and (\ref{h}) respectively). The form of the initial scattering data $a(\theta)$, $b(\theta)$ above constitute the main results of this section. It is clear that the Jost function $a(\theta)$ does not have singularities in the strip ${\rm Im}(\theta) \in [0,\pi]$, as it should. It is also important to note that it does not have zeroes either in that strip.

%%%%%%%%%%%%%%%%%%%%%%%%%%%%%%%%%%%%%%%%%%%%%%%%%%%%%%%%%%%%%%%%%%%%%%%
%%%%%%%%%%%%%%%%%%%%%%%%%%%%%%%%%%%%%%%%%%%%%%%%%%%%%%%%%%%%%%%%%%%%%%%

There is a useful check of these results, concerning the second Jost solution, defined by
\beqa
	\Psi_{-} &\stackrel{x\to\infty}=& c e^{\frc{ip_\theta x}2} \mato{c}1\\1\matf - de^{-\frc{ip_\theta x}2} \mato{c} 1\\-1 \matf\n
		&\stackrel{x\to-\infty}=& e^{\frc{-ip_\theta x}2} \mato{c} 1\\-1 \matf ~.\no
\eeqa
We see from the behaviour as $x\to -\infty$ that it is simply proportional to $\Psi_2$:
\beq
	\Psi_{-} = \frc{i\Psi_2}{g_- h_+}~.
\eeq
The coefficients are therefore
\beq
	c = \frc{i(1-\alpha)g_+}{g_-}=2ig_+=b(\theta+i\pi)~,\quad d=\frc{i(1-\beta) h_-}{h_+} = -2\pi i\alpha^2 h_-^2~.
\eeq
The various Wronskian relations obtained from the pairs $(\Psi_{+},\Psi_{-})$, $(\Psi_{+},\b\Psi_{-})$, $(\Psi_{+},\b\Psi_{+})$ are
\beq
	d=-a~,\quad |a|^2 + bc^* = 1~,\quad b^* =-b~.
\eeq
Our expressions for $a(\theta)$ and $b(\theta)$ may be readily shown to satisfy these requirements.
%%%%%%%%%%%%%%%%%%%%%%%%%%%%%%%%%%%%%%%%%%%%%%%%%%%%%%%%%%
%%%%%%%%%%%%%%%%%%%%%%%%%%%%%%%%%%%%%%%%%%%%%%%%%%%%%%%%%%
\section{The Gelfand-Levitan-Marchenko equations}\label{SecGLMOv}
%%%%%%%%%%%%%%%%%%%%%%%%%%%%%%%%%%%%%%%%%%%%%%%%%%%%%%%%%%
%%%%%%%%%%%%%%%%%%%%%%%%%%%%%%%%%%%%%%%%%%%%%%%%%%%%%%%%%%
In the previous section, we showed how the scattering data, $a(\lambda)$ and $b(\lambda ,t)$, are obtained, (\ref{scdata}) and (\ref{timeevol}) (with $\lambda=e^{\theta}$). In this section, we show how the solution to the sinh-Gordon PDE, $\varphi (x,t)$, is recovered from the scattering data. We follow the derivation of \cite{BernardBook}, although we use slightly different arguments in order to deal with a generically complex $\varphi(x,t)$.
%%%%%%%%%%%%%%%%%%%%%%%%%%%%%%%%%%%%%%%%%%%%%%%%%%%%%%%%%%
%%%%%%%%%%%%%%%%%%%%%%%%%%%%%%%%%%%%%%%%%%%%%%%%%%%%%%%%%%
\subsection{Fourier representations}
%%%%%%%%%%%%%%%%%%%%%%%%%%%%%%%%%%%%%%%%%%%%%%%%%%%%%%%%%%
%%%%%%%%%%%%%%%%%%%%%%%%%%%%%%%%%%%%%%%%%%%%%%%%%%%%%%%%%%
Throughout this sub-section we consider time-evolved quantities, but omit the explicit time dependences for ease of notation. The gauge transformed Jost solutions
$$
\Psi_{+}=e^{-\frac{\varphi}{2}\sigma_z}\widehat{\Psi_{+}}\quad{\rm and}\quad
\Psi_{-}=e^{-\frac{\varphi}{2}\sigma_z}\widehat{\Psi_{-}}
$$
have Fourier representation
\begin{align}
\widehat{\Psi_{+}}&=e^{ip_\lambda x/2}\left(\begin{array}{c}1 \\ 1\end{array} \right)
+\int_x^\infty \dd y\,\Big(\widehat{U_1}(x,y)+\lambda^{-1}\widehat{W_1}(x,y)\Big) e^{ip_\lambda y/2}\label{EqFourOne}\\
\widehat{\Psi_{-}}&=e^{-ip_\lambda x/2}\left(\begin{array}{c}1 \\ -1\end{array} \right)
+\int_{-\infty}^x \dd y\,\Big(\widehat{U_2}(x,y)+\lambda^{-1}\widehat{W_2}(x,y)\Big) e^{-ip_\lambda y/2}\label{EqFourTwo}\,,
\end{align}
where $\widehat{U_i}(x,y)$ and $\widehat{W_i}(x,y)$ are two component vector kernels. The gauge transformed linear equation satisfied by $\widehat{\Psi} = \widehat{\Psi_{+}}$ and $\widehat{\Psi} = \widehat{\Psi_{-}}$ is $(\partial_x-\widehat{A_x})\widehat{\Psi}=0$ with the connection
$$
\widehat{A_x}(x;\lambda)=i\left(\begin{array}{cc}
\frac{i}{2}(\partial_t\varphi-\partial_x\varphi) &  \frac{m}{4}\left(\lambda-\lambda^{-1}e^{2\varphi}\right)\\ 
\frac{m}{4}\left(\lambda-\lambda^{-1}e^{-2\varphi}\right)  & -\frac{i}{2}(\partial_t\varphi-\partial_x\varphi)
\end{array} 
\right)=\lambda\widehat{A_1}+\widehat{A_0}(x)+\lambda^{-1}\widehat{A_{-1}}(x)\;.
$$
Integrals of the type $\int_x^\infty dy f(y) e^{ip_\lambda y}$ are $O\lt(\lambda^{-1} e^{ip_\lambda x}\rt)$ as $\lambda\to\infty$, as can be seen from using the equality
$$
\lambda^{-2}e^{ip_\lambda y/2}=e^{ip_\lambda y/2}+\frac{4i}{m\lambda}\partial_{y}e^{ip_\lambda y/2}\,,
$$
and integrating by parts. Using this fact and that equality, we can equate terms of order $\lambda^{0}$ in the equation for $\widehat{\Psi_+}$ to obtain
$$
\left(1+\frac{4i}{m}\widehat{A_{1}}\right)\widehat{U_{1}}(x,x)=-\widehat{A_{0}}(x)\left(\begin{array}{c}1 \\ 1\end{array} \right)\,.
$$
This is equivalent to the following relation between derivatives of the field $\varphi(x,t)$ and the kernels:
\begin{equation}\label{EqnDerivs}
\left(\partial_{t}-\partial_{x}\right)\varphi(x)=2\left(\widehat{U_{1}}(x,x)^{+}-
\widehat{U_{1}}(x,x)^{-}\right)\,.
\end{equation}
In the above equation, $+$ and $-$ indicate, respectively, the first and second component of the vector. Further, by equating terms of order $\lambda^{-1}$ and substituting equation~(\ref{EqnDerivs}), we obtain
\begin{align}
e^{2\varphi(x)}&=1+\frac{4i}{m}\widehat{W_{1}}(x,x)^{-}-\frac{4i}{m}
\widehat{W_{1}}(x,x)^{+}
+\frac{16}{m^2}\left(\widehat{U_{1}}(x,x)^{-}-\widehat{U_{1}}(x,x)^{+}\right)
\widehat{U_{1}}(x,x)^{+}\nonumber\\
&\qquad-\frac{16}{m^2}\left(\partial_{x}\widehat{U_{1}}(x,y)^{+}+\partial_{y}
\widehat{U_{1}}(x,y)^{-}\right)|_{x=y}\,.
\label{EqnExp}
\end{align}
Note that this differs from formula 13.21 in~\cite{BernardBook}. Hence, knowing the kernels (which are functions of time through the time dependence of the Jost solutions), we may deduce the function $\varphi(x,t)$. The Gelfand-Levitan-Marchenko (GLM) equations express $\widehat{U_i}(x,y)$ and $\widehat{W_i}(x,y)$ in terms of the scattering data.
%%%%%%%%%%%%%%%%%%%%%%%%%%%%%%%%%%%%%%%%%%%%%%%%%%%%%%%%%%
%%%%%%%%%%%%%%%%%%%%%%%%%%%%%%%%%%%%%%%%%%%%%%%%%%%%%%%%%%
\subsection{The GLM equations}\label{SecGLM}
%%%%%%%%%%%%%%%%%%%%%%%%%%%%%%%%%%%%%%%%%%%%%%%%%%%%%%%%%%
%%%%%%%%%%%%%%%%%%%%%%%%%%%%%%%%%%%%%%%%%%%%%%%%%%%%%%%%%%
The previous subsection showed how the kernels are related to the solution, $\varphi (x,t)$, of the PDE. Here, we demonstrate how the kernels are obtained from the scattering data $a(\lambda)$ and $b(\lambda ,t)$ which we have obtained in section~\ref{SecScattDat}. The Jost solutions are related to their symmetry transformed versions by
\begin{align}
\Psi_{+}^\shift (\lambda,t)&=-\frac{b(-\lambda,t)}{a(\lambda)}\Psi_{+}(\lambda,t)+\frac{1}{a(\lambda)}\Psi_{-}(\lambda,t)\label{EqSym}\\
\Psi_{-}^\shift (\lambda,t)&=\frac{1}{a(\lambda)}\Psi_{+}(\lambda,t)+\frac{b(\lambda,t)}{a(\lambda)}\Psi_{-}(\lambda,t)\,.
\end{align}
The gauge transformed version of the first of these equations is
\begin{equation}\label{eqqq}
\frac{1}{a(\lambda)}\widehat{\Psi_{-}}(\lambda,t)=\widehat{\Psi_{+}^\shift}(\lambda,t)+\frac{b(-\lambda,t)}{a(\lambda)}\widehat{\Psi_{+}}(\lambda,t)~,
\end{equation}
into which we may substitute the Fourier representations of the gauge transformed Jost solutions, (\ref{EqFourOne}) and (\ref{EqFourTwo}). It is convenient to define
\beq
	r(\lambda ,t)=\frc{b(-\lambda ,t)}{a(\lambda)}~.
\eeq
Following the convention used in~\cite{BernardBook}, we will also make use of a definition of integrals on the real $\lambda$ line that avoids singularities:
\beq\label{presc}
\int_{-\infty}^{\infty}{\dd\lambda}=\lim_{\epsilon\to 0}\left(
\int_{-1/\epsilon}^{-\epsilon}{\dd\lambda}+\int_{\epsilon}^{1/\epsilon}{\dd\lambda}\right)~.
\eeq
Multiplying (\ref{eqqq}) by $\lambda^{j}e^{ip_\lambda y/2}$ for $j\in\{0,-1\}$ and integrating over $\lambda$ from $-\infty$ to $\infty$ with this prescription leads to the GLM equations:
\begin{align}
-\frac{2}{m}\sigma_z \widehat{U_1}(x,y,t)&=F_0(x+y,t)\left(\begin{array}{c}1 \\ 1\end{array} \right)
+\int_x^\infty \Big[F_0(y+z,t)\widehat{U_1}(x,z,t)+F_{-1}(y+z,t)\widehat{W_1}(x,z,t)\Big]\dd z\label{EqGLM1}\\
\frac{2}{m}\sigma_z \widehat{W_1}(x,y,t)&=F_{-1}(x+y,t)\left(\begin{array}{c}1 \\ 1\end{array} \right)
+\int_x^\infty \Big[F_{-1}(y+z,t)\widehat{U_1}(x,z,t)+F_{-2}(y+z,t)\widehat{W_1}(x,z,t)\Big]\dd z\label{EqGLM2}\,,
\end{align}
where use has been made of the following functions, for $j\in\{0,-1,-2\}$:
\begin{align}\label{EqFInt}
F_j(x,t)&=\frac{1}{4\pi}\int_{-\infty}^\infty \dd\lambda\, \lambda^j e^{ip_\lambda x/2}r(\lambda ,t)\no\\
&=\frac{1}{4\pi}\int_{-\infty}^\infty \dd\lambda\, \lambda^j e^{ip_\lambda x/2-iE_\lambda t}r(\lambda ,0)
\,.
\end{align}
These GLM equations are defined only for $y> x$, and the value for $y=x$ of their solution is defined by the limit $y\to x^+$.

%%%%%%%%%%%%%%%%%%%%%%%%%%%%%%%%%%%%%%%%%%%%%%%%%%%%%%%%%%
%%%%%%%%%%%%%%%%%%%%%%%%%%%%%%%%%%%%%%%%%%%%%%%%%%%%%%%%%%
\section{Asymptotic solutions to the GLM integral equations}
%%%%%%%%%%%%%%%%%%%%%%%%%%%%%%%%%%%%%%%%%%%%%%%%%%%%%%%%%%
%%%%%%%%%%%%%%%%%%%%%%%%%%%%%%%%%%%%%%%%%%%%%%%%%%%%%%%%%%
%%%%%%%%%%%%%%%%%%%%%%%%%%%%%%%%%%%%%%%%%%%%%%%%%%%%%%%%%%
%%%%%%%%%%%%%%%%%%%%%%%%%%%%%%%%%%%%%%%%%%%%%%%%%%%%%%%%%%
\subsection{The zero-time form factor expansion on the circle recovered}
%%%%%%%%%%%%%%%%%%%%%%%%%%%%%%%%%%%%%%%%%%%%%%%%%%%%%%%%%%
%%%%%%%%%%%%%%%%%%%%%%%%%%%%%%%%%%%%%%%%%%%%%%%%%%%%%%%%%%
In this section, we check that the machinery we have developed to solve the sinh-Gordon PDE for all time differences $t$, reproduces the known result for zero time difference. Recall that the functions $F_j (x,t)$ in the GLM equations are integrals of the scattering data, which are simply related to the one-particle finite-temperature form factors. However, the complete $t=0$ form factor expansion contains contributions from all particle numbers and rapidities. It is a non-trivial validation of our method, therefore, that the solution to the GLM equations contains terms in the $t=0$ expansion with more than one particle.

First, let us re-write the functions $F_j(x,0)$ as convergent sums instead of integrals. Since $a(\lambda)$ is analytic in the upper half plane, the poles in $r(\lambda ,0)$ contributing to the integral in equation~(\ref{EqFInt}) for $x>0$ are a result of the dependence on $b(-\lambda ,0)$. In terms of $\theta=\log(\lambda)$, the poles are at $\theta=i\pi/2+\theta_n$ for $n\in\Z$ (recall the notation (\ref{thetan})). The residues of the integrand at these poles are
\beq
\frac{T\exp\left(-mx\cosh(\theta_n)/2\right)}
{m\cosh(\theta_n)}\exp\left(2\int_{-\infty}^{\infty}
\frac{\dd\theta'}{2\pi}\frac{1}{\cosh(\theta_n-\theta')}
\log\left(\frac{1-e^{-m\cosh(\theta')/T}}{1+e^{-m\cosh(\theta')/T}}\right)\right)\,.
\eeq
Hence, the functions $F_j$ are given by the following sums
\begin{align*}
F_0(x,0)&=\sum_{n\in\Z}{-\frac{T e^{\theta_n-mx\cosh(\theta_n)/2}}{m\cosh(\theta_n)}
\exp\left(2\int_{-\infty}^{\infty}\frac{\dd\theta'}{2\pi}
\frac{1}{\cosh(\theta_n-\theta')}\log\left(\frac{1-e^{-m\cosh(\theta')/T}}{1+e^{-m\cosh(\theta')/T}}
\right)\right)}\\
F_{-1}(x,0)&=\sum_{n\in\Z}{i\frac{T e^{-mx\cosh(\theta_n)/2}}{m\cosh(\theta_n)}
\exp\left(2\int_{-\infty}^{\infty}\frac{\dd\theta'}{2\pi}
\frac{1}{\cosh(\theta_n-\theta')}\log\left(\frac{1-e^{-m\cosh(\theta')/T}}{1+e^{-m\cosh(\theta')/T}}
\right)\right)}\\
F_{-2}(x,0)&=\sum_{n\in\Z}{\frac{T e^{-\theta_n-mx\cosh(\theta_n)/2}}{m\cosh(\theta_n)}
\exp\left(2\int_{-\infty}^{\infty}\frac{\dd\theta'}{2\pi}
\frac{1}{\cosh(\theta_n-\theta')}\log\left(\frac{1-e^{-m\cosh(\theta')/T}}{1+e^{-m\cosh(\theta')/T}}
\right)\right)~.}
\end{align*}

We can obtain a large-distance expansion for $\varphi$ using the form factor expansion in the quantisation on the circle. With the functions
\beq
F_{N}(\theta_{n_1},\theta_{n_2},\ldots,\theta_{n_N})=\prod_{0<p<q\leq N}
\tanh\left({\frac{\theta_{n_p}-\theta_{n_q}}{2}}\right)\,,
\eeq
and
\beq
g(\theta)=\exp\left(\int_{-\infty}^{\infty}\frac{\dd\theta'}{2\pi}\frac{1}{\cosh(\theta-\theta')}
\log\left(\frac{1-\exp(-m\beta\cosh\theta')}{1+\exp(-m\beta\cosh\theta')}\right)\right)/\sqrt{m\beta\cosh\theta}~,
\eeq
the form factor expansion (\ref{circleffexp}) for the two-point correlation function of $\sigma$ operators is
\beqa\label{EqFF}
&& {}_\beta^{\hspace{-2mm}{\rm NS}}\bra\vac|\sigma(0,-ix)\sigma(0,0)|\vac\ket_\beta^{{\rm NS}} = \\ && \qquad\qquad
s_T^2 e^{-\Delta{\cal E} x}\sum_{N=0\atop N \ {\rm even}}^{\infty}\sum_{n_1,\ldots,n_N \in \Z}
\left(\prod_{i=1}^{N}
|g(\theta_{n_i})|^2\right)|F_{N}(\theta_{n_1},\theta_{n_2},\ldots,\theta_{n_N})|^2e^{-mx\sum_{i} \cosh(\theta_{n_i})}\,. \no
\eeqa
The same expansion, but containing only contributions from terms with odd particle numbers $N$, exists for the correlation function of $\mu$ operators. Let us gather terms into groups with fixed particle number and use a short-hand notation for them:
\beq
G(x,0) = \; {}_\beta^{\hspace{-2mm}{\rm NS}}\bra\vac|\sigma(0,-ix)\sigma(0,0)|\vac\ket_\beta^{{\rm NS}} =
s_T^2 e^{-\Delta{\cal E} x} \lt(1 + 2_{ff} + 4_{ff} + \ldots\rt)
\eeq
and
\beq
\t{G}(x,0) = \; {}_\beta^{\hspace{-2mm}{\rm NS}}\bra\vac|\mu(0,-ix)\mu(0,0)|\vac\ket_\beta^{{\rm NS}} =
s_T^2 e^{-\Delta{\cal E} x} \lt(1_{ff} + 3_{ff} + \ldots\rt)~.
\eeq
Then, $\varphi$ may be expanded in terms of form factors via its definition in equation~(\ref{EqGGtilde}) as
\beq
\varphi(x,0)=\tanh^{-1}\left(\frac{\tilde{G}(x,0)}{G(x,0)}\right)=2\lt[1_{ff}+\lt(\frac{1_{ff}^3}{3}-1_{ff}2_{ff}+3_{ff}\rt)+\ldots\rt]\label{EqnFFExpansion}\,.
\eeq
This is naturally a large distance expansion.

In the GLM equations, the functions $F_{j}(x,0)$  contain terms which are exponentially small at large distances $x$, as is clear from the sum expressions above. The iterative solution to the GLM equations, which is an expansion in powers of the number of functions $F_{j}(x,0)$, is therefore also an expansion of the type of the form factor expansion. It is natural to conjecture that the number of particles in the form factor expansion is in correspondance with the number of powers of $F_j(x,0)$ from the iterative solution of the GLM equations. Equation~(\ref{EqnExp}) may be written as
\begin{align}
\varphi(x,0)&=\frac{1}{2}\ln(1+1_{GLM}+2_{GLM}+3_{GLM}+\ldots)\nonumber\\
&=\frac{1_{GLM}}{2}+
\left(\frac{2_{GLM}}{2}-\frac{1_{GLM}^2}{4}\right)
+\left(\frac{3_{GLM}}{2}-\frac{1_{GLM}2_{GLM}}{2}+\frac{1_{GLM}^3}{6}\right)\label{EqnGLMExpansion}~,
\end{align}
in which $2_{GLM}$, for example, contains all terms present on the right hand side of equation~(\ref{EqnExp}) with quadratic powers of the functions $F_{j}(x,0)$. Our conjecture is that expansions~(\ref{EqnFFExpansion}) and~(\ref{EqnGLMExpansion}) should agree term by term. Let us show this explicitly for the linear term. The associated contributions to equation~(\ref{EqnExp}) containing only single powers of the functions $F_j (x,0)$ are
\beq
1_{GLM}=\frac{-4i}{m}\widehat{W_1}(x,x,0)^++\frac{4i}{m}\widehat{W_1}(x,x,0)^--\frac{16}{m^2}
\left(\partial_x\widehat{U_1}(x,y,0)^{+}+\partial_y\widehat{U_1}(x,y,0)^{-}\right)|_{y=x}\,,
\eeq
in which all terms in $\widehat{U}(x,y,0)$ and $\widehat{W}(x,y,0)$ are evaluated to one power of the functions $F_{j}(x,0)$. This is equivalent, therefore, to neglecting the contribution from the integrals, since these are quadratic or higher in powers of $F_j (x,0)$. We obtain
\begin{align}
\widehat{W_1}(x,x,0)^+&=\frac{iT}{2m}\sum_{n}\frac{e^{-m\cosh(\theta_{n})x}}
{\cosh(\theta_n)}
\exp\left(2\int_{-\infty}^{\infty}\frac{\dd\theta'}{2\pi}
\frac{1}{\cosh(\theta_n-\theta')}\log\left(\frac{1-e^{-m\cosh(\theta')/T}}{1+e^{-m\cosh(\theta')/T}}
\right)\right)\nonumber\\
\widehat{W_1}(x,x,0)^-&=-\widehat{W_1}(x,x,0)^+\nonumber\\
\widehat{U_1}(x,y,0)^+&=\frac{T}{2m}\sum_{n}e^{\theta_{n}}\frac{e^{-m\cosh(\theta_{n})(x+y)/2}}
{\cosh(\theta_n)}
\exp\left(2\int_{-\infty}^{\infty}\frac{\dd\theta'}{2\pi}
\frac{1}{\cosh(\theta_n-\theta')}\log\left(\frac{1-e^{-m\cosh(\theta')/T}}{1+e^{-m\cosh(\theta')/T}}
\right)\right)\nonumber\\
\widehat{U_1}(x,y,0)^-&=-\widehat{U_1}(x,y,0)^+\,.
\end{align}
Then, only the $\widehat{W_1}(x,x,0)$ terms contribute to $1_{GLM}$, since the contributions from
$\widehat{U_1}(x,y,0)^+$ and $\widehat{U_1}(x,y,0)^-$ cancel. We find
$$
1_{GLM}=\frac{4T}{m}\sum_{n}\frac{e^{-m\cosh(\theta_{n})x}}
{\cosh(\theta_n)}
\exp\left(2\int_{-\infty}^{\infty}\frac{\dd\theta'}{2\pi}
\frac{1}{\cosh(\theta_n-\theta')}\log\left(\frac{1-e^{-m\cosh(\theta')/T}}{1+e^{-m\cosh(\theta')/T}}
\right)\right)\,,
$$
in agreement with $1_{GLM}=4\ 1_{ff}$, with the identification $1/\beta=T$. We have also shown that $\left(\frac{2_{GLM}}{2}-\frac{1_{GLM}^2}{4}\right)=0$ by interating the integrals in the GLM equation once, and also that agreement is obtained for the cubic terms in the expansions. The three particle form factor, $3_{ff}$, contains the terms $|\tanh(\frac{\theta_{i}-\theta_{j}}{2})|^2$. The fact that the GLM equations recreate these terms is a non-trivial validation of the method, since the building blocks of the equations, $F_j (x,0)$ contain no such hyperbolic tangent functions themselves.
%%%%%%%%%%%%%%%%%%%%%%%%%%%%%%%%%%%%%%%%%%%%%%%%%%%%%%%%%%
%%%%%%%%%%%%%%%%%%%%%%%%%%%%%%%%%%%%%%%%%%%%%%%%%%%%%%%%%%
\subsection{Large-time asymptotics}\label{SecLarget}
%%%%%%%%%%%%%%%%%%%%%%%%%%%%%%%%%%%%%%%%%%%%%%%%%%%%%%%%%%
%%%%%%%%%%%%%%%%%%%%%%%%%%%%%%%%%%%%%%%%%%%%%%%%%%%%%%%%%%

We now seek a solution that is valid in the region $t+x\gg t-x,T^{-1},m^{-1}$, with $t>x>0$. Our solution will be valid for $m/T\neq0$ and $m (t-x)\neq0$. In that region, the residues of the integrand in equation~(\ref{EqFInt}) are no longer exponentially damped away from $\theta=0$; a new approach is required. The main problem is that we need to cross the light cone $x^2=t^2$. The choice of the region above, ``near'' the light cone, is for two reasons: it is the simplest region where we can illustrate how to cross the light cone in the GLM equations, and this region has not been investigated in the literature yet.

Notice that as $\lambda\to\infty$, we have $r(\lambda,0)\sim 2i$. Hence, from (\ref{EqFInt}), there is a delta-function contribution at $x=2t$ in $F_0(x,t)$. This is simple to extract, and we can write $F_0$ as a distribution:
$$
F_0(x,t) = \frac{2i}{m}\delta(x-2t) + F^P_0(x,t)~.
$$
The function $F^P_0(x,t)$ is the principal part of $F_0(x,t)$: it is equal to $F_0(x,t)$ for $|x-2t|>\ep$ and is zero (or any finite number) otherwise, where $\ep$ is some positive number that needs to be taken to zero after integration over $x$. Taking the principal part is important, because the asymptotics of $r(\lambda,0)$ also says that a pole will be present at $x=2t$. Hence, the integration over $x$ of an integrand that has $F^P_0(x,t)$ as a factor is the usual Cauchy principal value integral about that point. The delta-function contributes to the GLM equations in the time-like regime only, $t>x$. Let us consider the first GLM equation, (\ref{EqGLM1}). The kernel $\widehat{U_1}$ also acquires a delta-function piece, and we can also write it as a distribution:
$$
\widehat{U_1}(x,y)=\delta(x+y-2t)\left(\begin{array}{c}
-i \\ i
\end{array}\right) + \xi(x,y,t)\,,
$$
where that $\xi$ is a two-component vector that does not have delta-function contributions. It is expected to be itself a principal part of $\widehat{U_1}$ with respect to various points, including $x+y=2t$. The first GLM equation can then be written as (in the regime $t>x$):
\begin{align}\label{eqGLM1time}
-\frac{2}{m}&\sigma_z\xi(x,y,t)=F_0^P(x+y,t)\left(\begin{array}{c}1 \\ 1\end{array}\right) +F_0^P(y+2t-x,t)\left(\begin{array}{c}-i \\ i\end{array}\right) \nonumber\\
& + \int_x^{\infty} \Big[ F_0^P(y+z,t)\xi(x,z,t) + F_{-1}(y+z,t)\widehat{W_1}(x,z,t) \Big] \dd z
+ \frac{2i}{m}\xi(x,2t-y,t) \left(\begin{array}{c}1 \\ 1\end{array}\right) \Theta(2t-x-y)
\end{align}
where $\Theta$ is the step function. This should be seen as an equation for a distribution $\xi(x,y,t)$, and integrals as principal value integrals.

In appendix \ref{appintegrals}, we give a derivation of the asymptotic solution to these integral equations, up to the integral over $F_{-1}\widehat{W_1}$:
\beq\label{solxi}
\xi(x,y,t)=\frac{m}{2}F_0^P(x+y,t) \left(\begin{array}{c}-1 \\ 1\end{array}\right) ~.\quad
\eeq
Here we will just check that this function gives equality in (\ref{eqGLM1time}) up to the integral over $F_{-1}\widehat{W_1}$, and up to terms
\beq\label{negl}
	e^{-im\sqrt{t^2-(x+y)^2/4-i0^+}} O\lt((mt)^{-\infty}\rt)~, \quad t-x,~t-y~\mbox{fixed.}
\eeq
It will then be checked that the integral over $F_{-1}\widehat{W_1}$ gives also this type of term. The derivation given in appendix \ref{appintegrals} starts from a more general and natural ansatz and shows that the solution (\ref{solxi}) is the only one possible. We will not go into discussion of the uniqueness of the solution to the GLM equations; uniqueness proofs can be found, for instance, in \cite{FadTakBook} for the sine-Gordon model, but we do not know if uniqueness has been assessed in our case, where $\varphi$ has a complex region. Naturally, the solution for $\varphi$ with appropriate initial condition at $t=0$ is unique (with the extra condition that $\varphi(x,t) = \varphi(x,t-i0^+)$ in order to go through the singularity on the light-cone $x^2=t^2$), and our solution to the GLM equations should give it.

First, in the region $x/2+t\gg x/2-t$, the integrals in equation~(\ref{EqFInt}) are dominated by contributions from large $\theta$ (i.e. as $\lambda\to\pm\infty$). We can then expand $r(\theta,0)$ and $r(\theta+i\pi,0)$ in powers of $e^{-\theta}$. The integrals for $F_j(x,t)$ may be done analytically on each term of the expansion, and give an expansion in that region. We have
\begin{align}
r(\theta ,0)& =4\pi g_+(\theta)h_+(\theta)^2 =2i\sum_{\mu,\nu =0}^\infty{c_{\mu\nu}(T)e^{-\mu \theta}e^{-\nu E_\theta/T}}\\
r(\theta+i\pi ,0)&=-4\pi g_-(\theta)h_-(\theta)^2=2i\sum_{\mu,\nu=0}^\infty{\tilde{c}_{\mu\nu}(T)e^{-\mu \theta}e^{-\nu E_\theta/T}}\,,
\end{align}
with coefficients, $c_{\mu\nu}(T),\t{c}_{\mu\nu}(T)$, coming from the expansion of the functions (\ref{g}) and (\ref{h}) in powers of $e^{-\theta}$. They are, in general, integral expressions with dependence only on temperature (and mass). Note, however, that $c_{00}=1$ and $\tilde{c}_{\mu 0}=0$ for all $\mu$. With use of the identity
$$
\int_{0}^{\infty}{u^c e^{-au-b/u}\dd u}=2\left(\frac{b}{a}\right)^{\frac{1+c}{2}}
K_{-1-c}(2\sqrt{ab})\,,
$$
the above form of $r(\theta ,t)$ leads to the following expression for the functions $F_j (x,t)$:
\begin{align}
F_j&(x,t)=\frac{i}{\pi}\sum_{\mu,\nu=0}^\infty{c_{\mu\nu}(T)\left(\frac{it-ix/2+\nu/T}{it+ix/2+\nu/T}\right)^{\frac{\mu-1-j}{2}}
K_{1+j-\mu}(m\sqrt{(it+ix/2+\nu/T)(it-ix/2+\nu/T)})}\nonumber\\
&\quad +\frac{i}{\pi}\sum_{\mu,\nu=0}^\infty{\tilde{c}_{\mu\nu}(T)\left(\frac{-it+ix/2+\nu/T}{-it-ix/2+\nu/T}\right)^{\frac{\mu-1-j}{2}}
K_{1+j-\mu}(m\sqrt{(-it-ix/2+\nu/T)(-it+ix/2+\nu/T)})}\,,\label{expFj}
\end{align}
where $K_s(x)$ is the modified Bessel function of the second kind. Above, we take the principal branch of the square-root and in the cases where $\nu=0$, we consider $\nu=0^+$. For $j=0$, the above gives $F_0^P(x,t)$ for $x\neq 2t$. Note that the term with $\mu=\nu=0$ in the first series gives the expected pole as $x\to 2t$.

In order to obtain terms up to (\ref{negl}), we disregard all terms with $\nu\neq0$; in particular, the second series above can be neglected. Now we see that on the right-hand side of (\ref{eqGLM1time}), the integration over $F_0^P(y+z,t)\xi(x,z,t)$ has one or two poles about which the principal value has to be taken: one at $z=2t-y$ if $y<2t-x$, and one at $z=2t-x$ coming from $\xi(x,z,t)$. These principal values can be evaluated by closing the $z$-contour around these points, and substracting the resulting half-residue. Closing the contour can only be made in the upper-half $z$-plane, because of the branch structure of the argument of the modified Bessel functions in the first series in the expansion for $F_0$ above. Doing this exactly cancels the terms $F_0^P(y+2t-x,t) (-i,i)^T$ and $\frac{2i}{m}\xi(x,2t-y,t)(1,1)^T \Theta(2t-x-y)$ that were coming from the delta-functions. We are left with an integral of the form
\beq\label{soi}
	\int_{{\cal C}_x} F_0(y+z,t)F_0(x+z,t) \dd z
\eeq
where the contour ${{\cal C}_x}$ starts at $x$ and finishes at $\infty$ by going through the upper-half plane. Everywhere on this contour, the leading asymptotic $K_{1-\mu}(u) \sim \sqrt{\frc{\pi}{2u}} e^{-u}$ of the modified Bessel functions at large argument can be replaced, and the resulting integrals are convergent. These integrals are of the type (\ref{negl}), so can be neglected. More precisely, they are $\exp[-im\sqrt{t^2-x^2} -im \sqrt{t^2-(x+y)^2/4}] O(t^a)$ for some power $a$. For the case $\mu=0$, there is a logarithmic singularity as $y\to 2t-x$, but no other singularities occur. Hence the neglected terms in $\xi(x,y,t)$ can have at most logarithmic singularities, which are integrable, so under integration in (\ref{eqGLM1time}) they give again terms of the same type (\ref{negl}).

Hence, we have
\beq
	\widehat{U_1}(x,y,t) = \frc{m}2 F_0(x+y,t) \mato{c} -1\\1 \matf + \mbox{ terms of the type (\ref{negl}).}
\eeq
This gives no contribution to $\varphi (x,t)$ to the order of interest in equation~(\ref{EqnExp}), since it is a function of $x+y$ and proportional to $(1,-1)^T$, and the quadratic terms in equation~(\ref{EqnExp}) can be neglected. On substitution into the GLM equation for $\widehat{W_1}$, excluding the term corresponding to the integral over $F_{-2}\widehat{W_1}$, this result leads to:
\beq
\frac{2}{m}\sigma_z\widehat{W_1}(x,y,t)=F_{-1}(x+y,t)\mato{c} 1\\1\matf + \int_x^\infty F_{-1}(y+z,t) F_0(x+z,t) \dd z
\label{EqWs}\,.
\eeq
Again, the integral part gives no contribution to the leading terms (since the delta-function piece is cancelled by closing the contour of the principal value integral in the upper half plane), so that we have
\begin{equation}
\widehat{W_1}(x,y,t) = \frc{m}2 F_{-1}(x+y,t) \mato{c} 1\\-1\matf + \mbox{ terms of the type (\ref{negl}).}
\end{equation}

Recall that we have as yet ignored two of the integrals in the GLM equations: $\int F_{-1}\widehat{W_1}$ and $\int F_{-2}\widehat{W_1}$. The results above vindicate this omission; such integrals lead to terms of the type (\ref{negl}).

Note finally that the space-like region $x+t\gg x-t$ with $x>t>0$ can also be investigated with similar techniques, with the simplification that no delta-function contribution appears and that no singularities have to be dealt with. The results for $\widehat{U_1}$ and $\widehat{W_1}$ above are still valid in that region.

%%%%%%%%%%%%%%%%%%%%%%%%%%%%%%%%%%%%%%%%%%%%%%%%%%%%%%%%%%
%%%%%%%%%%%%%%%%%%%%%%%%%%%%%%%%%%%%%%%%%%%%%%%%%%%%%%%%%%
\section{Results}
%%%%%%%%%%%%%%%%%%%%%%%%%%%%%%%%%%%%%%%%%%%%%%%%%%%%%%%%%%
%%%%%%%%%%%%%%%%%%%%%%%%%%%%%%%%%%%%%%%%%%%%%%%%%%%%%%%%%%

All our results are valid for $|t-x|m\neq0$ and $m/T\neq0$. The results of the previous section can be combined with (\ref{EqnExp}) to give
\beq
\varphi(x,t) = -2i F_{-1}(2x,t) + e^{-im\sqrt{t^2-x^2}} O\lt((mt)^{-\infty}\rt)~, \quad t-x~\mbox{fixed.}
\eeq
That is,
\begin{equation}\label{expphi}
\varphi (x,t)=\frac{2}{\pi}
\sum_{\mu=0}^\infty c_{\mu}(T)
\left(\frac{t-x}{t+x}\right)^{\frac{\mu}{2}} K_\mu(im\sqrt{t^2-x^2}) + e^{-im\sqrt{t^2-x^2}} O\lt((mt)^{-\infty}\rt)~, \quad t-x~\mbox{fixed,}
\end{equation}
where $c_\mu(T) \equiv c_{\mu0}(T)$. Recall that the coefficients $c_{\mu0}(T)$ come from the expansion of the one-particle finite-temperature form factors in powers of $e^{-\theta}$. They may be calculated as integrals over rapidites, and are given by the generating function
\beq
	\sum_{\mu=0}^\infty c_{\mu}(T) q^\mu = \exp\lt[-\frc{2i}\pi \sum_{\omega=0}^{\infty} q^{2\omega+1} \int_{-\infty-i0^+}^{\infty-i0^+} \dd\theta
	e^{(2\omega+1)\theta} \ln\left(\frac{1+e^{-m\cosh(\theta)/T}}{1-e^{-m\cosh(\theta)/T}} \right)\rt]~.
\eeq

The expansion (\ref{expphi}) is a solution to the linearized version of the sinh-Gordon equation (\ref{shG}) (that is, the Klein-Gordon equation). For convenience, we will use here the variables
\[
	v = t-x~,\quad w=t+x~.
\]
In fact, (\ref{expphi}) is an infinite linear combination of solutions $\Phi_\mu(v,w) = (v/w)^{\frac{\mu}{2}} K_\mu(im\sqrt{vw})$ with spins $\mu=0,1,2,\ldots$ and temperature-dependent coefficients $c_\mu(T)$.

In order to obtain the correlation functions, we need to evaluate $\chi$ through (\ref{eqChi}), (\ref{eqOthers}). These equations for $\chi$ leave room for three undetermined, generically temperature-dependent constants:
\beq
	\chi = -(A+Bx+Ct) + \t\chi
\eeq
with $\lim_{w\to\infty} \t\chi$ = 0. At $T=0$, the constants $A,B,C$ all vanish. In general, they cannot be determined using the integrable PDEs for correlation functions; an alternative method would be required. We expect that $B+C>0$, which represents a decay of the thermal correlation functions at large time in the region we are looking at.

The leading terms for $\t\chi$ are obtained by keeping only the quadratic power of $\varphi$ in (\ref{eqChi}). It is a non-trivial fact that any solution $\Phi_\mu(v,w)$ to the Klein-Gordon equation gives rise, through (\ref{eqChi}) and (\ref{eqOthers}), to a consistent large-$w$ expansion for $\t\chi$ of the form $e^{-2im\sqrt{vw}}$ times a series in growing integer powers of $w^{-\frc12}$.

The large-$w$ expansions of $\varphi$ and $\t\chi$ can be written
\beqa\label{expphi2}
	\varphi &=& \sqrt{\frc{-2i}{\pi m}} (vw)^{-\frc14} e^{-im\sqrt{vw}} \lt(1+\frc{g_1(v)}{m\sqrt{vw}} + \frc{g_2(v)}{m^2vw} + \frc{g_3(v)}{m^3(vw)^{\frc32}}
	+ O((mw)^{-2})\rt) \\
	\t\chi &=& \frc{i}{2\pi m} (vw)^{-\frc12} e^{-2im\sqrt{vw}} \lt(1+\frc{f_1(v)}{m\sqrt{vw}} + \frc{f_2(v)}{m^2vw} + \frc{f_3(v)}{m^3(vw)^{\frc32}} + O((mw)^{-2})\rt)~.
\eeqa
From (\ref{expphi}), we immediately have
\beqa
	g_1(v) &=& \frc{i}8 + c_1 mv \n
	g_2(v) &=& -\frc{9}{128} - \frc{3ic_1}{8} mv + c_2 (mv)^2 \\
	g_3(v) &=& -\frc{75 i}{1024} + \frc{15 c_1}{128} mv - \frc{15 i c_2}{8} (mv)^2 + c_3 (mv)^3~. \no
\eeqa
Then, solving (\ref{eqChi}) to quadratic order of $\varphi$, or solving (\ref{eqOthers}), we find consistently
\beqa
	f_1(v) &=& \frc{3i}4 + 2c_1 mv \n
	f_2(v) &=& -\frc{33}{32} + \frc{i c_1}2 mv + (2c_2+c_1^2) (mv)^2 \\
	f_3(v) &=& -\frc{255 i}{128} - \frc{9 c_1}{16} mv - \frc{i}4 (c_1^2+10c_2) (mv)^2 + 2(c_1c_2 + c_3) (mv)^3~. \no
\eeqa
The correlation functions (\ref{EqGGtilde}) are obtained from these series expansion:
\beqa
	G &=& s_T^2 e^{-A-Bx -Ct}\lt(1+\frc{\t\chi}2 + \frc{\varphi^2}8\rt) + e^{-im\sqrt{vw}} O((mw)^{-\infty})\n
	\t{G} &=& s_T^2 e^{-A-Bx -Ct}\frc{\varphi}2 + e^{-im\sqrt{vw}} O((mw)^{-\infty})~. \label{tGphi}
\eeqa
For $G$ this gives:
\beqa
	G &=& s_T^2 e^{-A-Bx -Ct}\Bigg[1  +  e^{-2im\sqrt{vw}} \times \\ && \times \lt(- \frc1{8\pi m^2 vw} - \frc{7i + 8c_1 mv}{32 \pi m^3 (vw)^{\frc32}} +
	\frc{117 -48ic_1mv - 32(c_1^2+2c_2)(mv)^2}{256\pi m^4(vw)^2} + O((mw)^{-\frc52}) \rt) \Bigg]~. \no
\eeqa
It is interesting to see that the first two terms inside the parenthesis are temperature-independent. Taking $T=0$ (where $A=B=C=0$), they can be seen to agree with the standard zero-temperature form factor expansion, which is valid also in the time-like region. The next terms acquire temperature corrections. Note that it was necessary to calculate $g_3$ and $f_3$ in order to obtain the last term, although it only involves the constants $c_1$ and $c_2$.

Note also that the same series expansion holds in the space-like regime $x+t\gg x-t,T^{-1},m^{-1}$ with $x>t>0$. In this case, we need to take $v$ with a phase $e^{-i\pi}$. Since there is a path from $v>0$ to $v<0$ through the negative imaginary $v$-plane on which the series expansion is valid at every point, we expect that the constants $A,B,C$ are the same in this space-like regime.

From (\ref{tGphi}), the expansion (\ref{expphi2}) can be used directly to give an expansion for $\t{G}$, valid for $x+t\gg |x-t|,T^{-1},m^{-1}$ with $x-t$ positive or negative (but, again, non-zero). In fact, the formula
\[
	\t{G} = -i s_T^2 e^{-A-Bx-Ct} \lt(F_{-1}(2x,t) + e^{-im\sqrt{t^2-x^2}} O\lt((mt)^{-\infty}\rt) \rt)~,  \quad t-x~\mbox{fixed}
\]
shows that it is, up to the pre-factor $e^{-A-Bx-Ct}$, just the finite-temperature form-factor expansion up to one particle. The one-particle finite-temperature form factor expansion, initially an expansion at large distances, can indeed be continued to a large-time expansion without problems (in particular, to the region we are looking at here). Of course, the higher-particle contributions, under such continuation, would give contributions at the ``one-particle'' order as well, because of the poles in rapidity space (as briefly discussed in section \ref{sectftcf}). Our results show that in our region of interest, these contributions only enter the exponential pre-factor, determining the constants $A,B,C$. In fact, the constant $C$ is certainly expected to be modified in going from the large-distance expansion to the expansion in the region we are looking at. Indeed, in a large-distance expansion it is $0$, but there is no reason to believe that it should be zero here.

%%%%%%%%%%%%%%%%%%%%%%%%%%%%%%%%%%%%%%%%%%%%%%%%%%%%%%%%%%
%%%%%%%%%%%%%%%%%%%%%%%%%%%%%%%%%%%%%%%%%%%%%%%%%%%%%%%%%%
\section{Conclusion}
%%%%%%%%%%%%%%%%%%%%%%%%%%%%%%%%%%%%%%%%%%%%%%%%%%%%%%%%%%
%%%%%%%%%%%%%%%%%%%%%%%%%%%%%%%%%%%%%%%%%%%%%%%%%%%%%%%%%%
Inspired by the link with finite-temperature correlation functions of spin operators in the transverse Ising model near its quantum critical point, we have investigated the dynamical two-point correlation functions of twist fields in the free Majorana theory at finite temperature and real time. These objects are hard to calculate by the well-known mapping to the vacuum theory on the circle, since this approach leads naturally to imaginary time and the analytic continuation to real time is generically plagued by singularites. Instead, we used the fact that the correlation functions satisfy a well-known set of integrable partial differential equations (PDEs), including the sinh-Gordon equation. These are amenable to solution by the classical inverse scattering method, which is a generalisation of the Fourier transform to tackle such nonlinear, integrable PDEs. The method relies on solving the associated scattering problem and mapping the results back to the solution of the PDEs. The main objects are the scattering data, related to scattering solutions with incoming plane wave boundary conditions at $\pm\infty$. The scattering data evolve in a simple way with time, and an important step is to determine the initial scattering data.

We evaluated the initial scattering data that correspond to the correlation functions at finite temperature using the Liouville space of QFT (Hilbert space of finite-temperature states), and found that they are simply related to the finite-temperature form factors of the twist fields. Following the classical inverse scattering method, the solution to the PDEs are then obtained from the solution to a pair of coupled Volterra linear integral equations (the GLM equations). These equations (\ref{EqGLM1},\ref{EqGLM2}), with the kernels (\ref{EqFInt}) containing the initial scattering data, and the relation to the sinh-Gordon solution (\ref{EqnExp}), are part of our main results. We showed that these integral equations reproduce the known results for finite-temperature correlation functions at zero time (the form factor expansion in the quantisation on the circle) term by term for the first few terms. From the GLM equations, we then evaluated the expansion of the correlation functions at large times, in the region where the space variable is kept of the order of the time variable. To our knowledge, this region, ``near'' the light-cone, had not been investigated before in the literature. We found that the one-particle contributions to the finite-temperature form factor expansion, which naturally gives an expansion at finite time and large distances, also gives the right expansion in that region, up to exponentially smaller terms, and up to a factor of an exponential in space and time which is beyond the reach of the present method. This is non-trivial, since higher-particle terms in the form factor expansion are expected to give contributions to the same order as that of the one-particle terms in that region, due to singularities in the rapidity variables. Hence we found that they can only contribute to the exponential factor.

It would be very interesting to use these techniques to obtain the large-time expansion in different regions, or to solve numerically the linear integral equations in order to obtain correlation functions at all times. This would clarify the various approximate techniques that have been used up to now for studying finite-temperature Ising correlation functions, some of them being also applied to interacting integrable models. Our methods can be applied whenever there exists an integrable PDE describing correlation functions (as is known to happen for twist fields in free fermionic models), and applying it to other models may further clarify the structure of finite-temperature correlation functions. Also, it could be instructive for the theory of integrable PDEs to further investigate the connection between massive free models of QFT and integrable PDEs, and in particular, between the form factors and solutions to the GLM equations. In particular, it would be interesting to understand further the origin of the complex region of the sinh-Gordon solution from the viewpoint of the PDE. For instance, it seems, although it is not entirely clear, that the information about how to go through the light-cone was already included in the initial scattering data, since the derivation of our solution did not make use of any asumption for crossing the light-cone. In a related direction, we wish to point out that, in a similar fashion to the zero-temperature case and the relation to Painlev\'e equations, QFT ideas and results can give rise to non-trivial conjectures concerning solutions to integrable PDEs in the finite-temperature case. Finally, of course, our method cannot be directly applied to correlation functions in more complicated interacting integrable models, since there is generically no PDE description. However, it may be worth looking for a generalisation of the GLM equations that give rise to form factor expansions at zero temperature in interacting integrable models; that may open the way to finite-temperature form factors in such models, to new ways of generating form factors, and to new representations for correlation functions.
\bigskip

{\bf Acknowledgments}

The authors wish to thank P. Bowcock, J. Cardy, P. Dorey and F. Essler for useful discussions and comments on the manuscript. B.D. acknowledges partial support by an EPSRC post-doctoral fellowship, grant GR/S91086/01, and the Rudolf Peierls Centre for Theoretical Physics at Oxford University, where the fellowship was held and where a large part of this work was carried out. A.G. acknowledges support from an EPSRC Studentship.

%%%%%%%%%%%%%%%%%%%%%%%%%%%%%%%%%%%%%%%%%%%%%%%%%%%%%%%%%%
\appendix
%%%%%%%%%%%%%%%%%%%%%%%%%%%%%%%%%%%%%%%%%%%%%%%%%%%%%%%%%%
%%%%%%%%%%%%%%%%%%%%%%%%%%%%%%%%%%%%%%%%%%%%%%%%%%%%%%%%%%
\section{Derivation of the symmetric solution}\label{SecDerivSp}
%%%%%%%%%%%%%%%%%%%%%%%%%%%%%%%%%%%%%%%%%%%%%%%%%%%%%%%%%%
%%%%%%%%%%%%%%%%%%%%%%%%%%%%%%%%%%%%%%%%%%%%%%%%%%%%%%%%%%

In this appendix, we show that certain objects in the QFT model satisfy the linear problem associated to the sinh-Gordon differential equation. Following~\cite{FonsecaZ03}, let us consider the doubled theory: two copies (labelled by $a$ and $b$) of the Majorana fermion, with anticommuting fundamental fields and factorizing correlation functions. Symmetries in the doubled theory lead to trace identities for correlation functions incorporating the conserved charge. For example, the following integral is conserved:
\begin{align*}
Y_1&=\frac{1}{2\pi}\int_{-\infty}^{\infty}
\left(\psi_a(\xx) \p_s\psi_b(\xx)+i\frac{m}{2}\bar{\psi}_a(\xx)\psi_b(\xx)\right)\dd x\\
&=\frac{m}{2}\int_{-\infty}^{\infty}e^\beta \left(A(\beta)^{\dagger}B(\beta)+B(\beta)^{\dagger}A(\beta)\right)\frac{\dd\beta}{2\pi}~,
\end{align*}
where $A(\beta),B(\beta)$ are the annihilation operators of particles with rapidity $\beta$ for the copies $a,b$ respectively (their hermitian conjugates are the corresponding creation operators) and $\psi_a,\psi_b$ are the fundamental fermionic fields for copies $a,b$. The symbol $\xx$ represents the space-time doublet $(x,t)$, and the derivative $\p_s=(\p_x-\p_t)/2$ is the ``holomorphic'' derivative $\p$ that defined after (\ref{edm}). Therefore, the following trace identity holds:
\beq
\lt\bra [Y_1,\sigma_a(\xx) \sigma_b(\xx) A^{\dagger}(\beta')\sigma_a(\xx')\mu_b(\xx')]\rt\ket_T=0~.
\eeq
The twist fields $\sigma(\xx)$ are bosonic, whilst the $\mu(\xx)$ are fermionic. The commutator of the fundamental field with twist fields may be derived from the definition of twist fields as highest weight states, see~\cite{FonsecaZ03}. Those of interest to us are
\begin{align}
[Y_1,\sigma_{a}(\xx)\sigma_{b}(\xx)]&=-\p_s\mu_{a}(\xx)\mu_{b}(x)+\mu_{a}(\xx)\p_s\mu_{b}(\xx)\nonumber\\
[Y_1,\sigma_{a}(\xx)\mu_{b}(\xx)]&=-i\p_s\sigma_{a}(\xx)\mu_{b}(x)+i\sigma_{a}(\xx)\p_s\mu_{b}(\xx)~.\label{eq:app_corr}
\end{align}
The trace identity defined above is therefore equivalent to
\begin{align}
\big\bra
&\left[-\left(\p_s\mu_a(\xx)\right) \mu_b(\xx)+\mu_a(\xx) \p_s\mu_b(\xx)\right] A^{\dagger} (\beta')\sigma_a(\xx')\mu_b(\xx')\big\ket_T\no\\
&+\big\bra\sigma_a(\xx) \sigma_b(\xx) \left[
\frac{m}{2}\int_{-\infty}^{\infty}e^\beta (A(\beta)^{\dagger}B(\beta)+B(\beta)^{\dagger}A({\beta}))\frac{\dd\beta}{2\pi},
A^{\dagger} (\beta')\right]\sigma_a(\xx')\mu_b(\xx')\big\ket_T\no\\
&+\big\bra\sigma_a(\xx) \sigma_b(\xx) A^{\dagger} (\beta')
\left[i\left(\p_{s'}\mu_a(\xx')\right)\sigma_b(\xx')-i\mu_a(\xx')
\p_{s'} \sigma_b(\xx')\right]\big\ket_{T}=0~.
\end{align}
The correlation functions involving a product fields in both copies of the theory factorize. We define
\begin{align*}
E(\xx+\xx',\beta) \tilde{F}(\xx-\xx',\beta)\equiv\bra \mu(\xx) A^{\dagger}(\beta) \sigma(\xx')\ket_T\\
E(\xx+\xx',\beta) F(\xx-\xx',\beta)\equiv\bra \sigma(\xx) A^{\dagger}(\beta) \mu(\xx')\ket_T~.
\end{align*}
$E(\xx,\beta)$ are the centre of mass plane waves $e^{ip_\beta x/2 -iE_\beta t/2}$ and the functions $F$ are the finite temperature traces defined in section~\ref{SecSpSol}. With these definitions, the above equation becomes
\begin{align*}
&\p_s (E(\xx+\xx')\tilde{F}(\xx-\xx'))\tilde{G}(\xx-\xx')-E(\xx+\xx')\tilde{F}(\xx-\xx')\p_s\tilde{G}(\xx-\xx')\\
&+\sigma_a(\xx)\sigma_b(\xx)\frac{m}{2}\int e^{\beta} B(\beta)^{\dagger}
\delta(\beta-\beta')\sigma_a(\xx')\mu_b(\xx')\dd\beta\\
&+i\p_{s'}(E(\xx+\xx')F(\xx-\xx'))G(\xx-\xx')-iE(\xx+\xx')F(\xx-\xx')\p_{s'}G(\xx-\xx')=0~,
\end{align*}
where $G,\tilde{G}$ are the correlation function of twist fields, as in the text. This expression may be simplified to (dropping the functional dependencies for clarity)
\beq
\p_s (E\tilde{F})\tilde{G}-E\tilde{F}\p_s \tilde{G}+GEF \frac{me^{\beta'}}{2}
+i\p_{s'}(EF)G-iEF\p_{s'}G=0~.
\eeq
The trace of the same conserved charge with $\sigma_a(\xx) \mu_b(\xx) A^{\dagger}\sigma_a(\xx)\sigma_b(\xx')$ leads to a similar differential equation
\beq
i\p_s (E\tilde{F})G-iE\tilde{F}\p_s G+GE\tilde{F} \frac{me^{\beta'}}{2}
-\p_{s'}(EF)\tilde{G}+EF\p_{s'}\tilde{G}=0~.
\eeq
Using:
\begin{align}
\p_s G &= \frac{\p_s\chi}{2}G+\frac{\p_s\varphi}{2}\tilde{G}\no\\
\p_s \tilde{G} &= \frac{\p_s\chi}{2}\tilde{G}+\frac{\p_s\varphi}{2}G~,
\end{align}
and also $\p_{s'}G=-\p_s G$, and $\p_s E=\p_{s'} E=ime^{\beta}E/4$, and defining $f=e^{-\chi/2}F$, we find
\begin{align}
\p_s(f+i\tilde{f})=-i\frac{me^{\beta-\varphi}}{4}(f-i\tilde{f})+\frac{\p_s\varphi}{2}(f+i\tilde{f})\no\\
\p_s(f-i\tilde{f})=-i\frac{me^{\beta+\varphi}}{4}(f+i\tilde{f})-\frac{\p_s\varphi}{2}(f-i\tilde{f})~.\label{eq:app_one}
\end{align}
A related set of equations are obtained from the conserved charge:
\begin{align*}
Y_{-1}&=\frac{1}{2\pi}\int_{-\infty}^{\infty}
\left(\bar{\psi}_a(\xx)\bar{\p}_s\bar{\psi}_b(\xx)-i\frac{m}{2}\bar{\psi}_b(\xx)\psi_a(\xx)\right)\dd {x}\\
&=\frac{m}{2}\int_{-\infty}^{\infty}e^{-\beta} \left(A(\beta)^{\dagger}B(\beta)+B(\beta)^{\dagger}A(\beta)\right)\frac{\dd\beta}{2\pi}
\end{align*}
where $\b{\p}_s = (\p_x+\p_t)/2$. A pair of relations like those in equation~(\ref{eq:app_corr}) for $Y_{-1}$, but with the replacements $\p_s\rightarrow\bar{\p}_s$ and $i\rightarrow -i$ on the right hand side, is used to obtain the following relations:
\begin{align}
\bar{\p}_s(f+i\tilde{f})=i\frac{me^{-\beta+\varphi}}{4}(f-i\tilde{f})-\frac{\bar{\p}_s\varphi}{2}(f+i\tilde{f})\no\\
\bar{\p}_s(f-i\tilde{f})=i\frac{me^{-\beta-\varphi}}{4}(f+i\tilde{f})+\frac{\bar{\p}_s\varphi}{2}(f-i\tilde{f})~.\label{eq:app_two}
\end{align}
Using $\p_{x}=\p_s+\bar{\p}_s$, a linear combination of equations~(\ref{eq:app_one})~and~(\ref{eq:app_two}) shows that the vector $\Psi=(\tilde{f}-if,\tilde{f}+if)$ satisfies the equation $(\p_{x}-A_{x})\Psi=0$, where $A_x$ is the matrix associated to the zero-curvature formulation of the sinh-Gordon equation, as claimed in the text. It is referred to as the `special' solution.

%%%%%%%%%%%%%%%%%%%%%%%%%%%%%%%%%%%%%%%%%%%%%%%%%%%%%%%%%%
%%%%%%%%%%%%%%%%%%%%%%%%%%%%%%%%%%%%%%%%%%%%%%%%%%%%%%%%%%
\section{Asymptotics for large and small $\lambda$}\label{SecAsym}
%%%%%%%%%%%%%%%%%%%%%%%%%%%%%%%%%%%%%%%%%%%%%%%%%%%%%%%%%%
%%%%%%%%%%%%%%%%%%%%%%%%%%%%%%%%%%%%%%%%%%%%%%%%%%%%%%%%%%
Here, we show how to extract the behaviour of $a(\lambda)$ and $b(\lambda ,0)$ as $\theta\to\pm\infty$ by demonstrating the form of the Jost solution in these limits. This is one of the constraints on the form of $\alpha(\lambda)$ in section~\ref{SecScattDat}. We find that the Jost solutions $\Psi_{+}$ and $\Psi_{-}$ have the following $\lambda$ asymptotics for all $x$:
\begin{align}
\Psi_{+}&=e^{ip_\lambda x/2}\left( e^{-\varphi\sigma_z /2} \left(\begin{array}{c}1\\ 1
\end{array}  \right)+O\left(\frac{1}{|\lambda|}\right)\right),\quad |\lambda|\rightarrow\infty \label{EqLambAsym}\\
\Psi_{-}&=e^{-ip_\lambda x/2}\left( e^{-\varphi\sigma_z /2} \left(\begin{array}{c}1\\ -1
\end{array}  \right)+O\left(\frac{1}{|\lambda|}\right)\right),\quad |\lambda|\rightarrow\infty \\
\Psi_{+}&=e^{ip_\lambda x/2}\left( e^{\varphi\sigma_z /2} \left(\begin{array}{c}1\\ 1
\end{array}  \right)+O\left(|\lambda|\right)\right),\quad |\lambda|\rightarrow 0 \\
\Psi_{-}&=e^{-ip_\lambda x/2}\left( e^{\varphi\sigma_z /2} \left(\begin{array}{c}1\\ -1
\end{array}  \right)+O\left(|\lambda|\right)\right),\quad |\lambda|\rightarrow 0
\end{align}

The derivation of the first equation proceeds as follows. Start from the gauge transformed function
$$
\Psi_{+}=e^{ip_\lambda x/2-\varphi\sigma_z/2}\widehat{\Psi_{+}}\,,
$$
which satisfies the linear equation $[\partial_x+ip_\lambda(1-\sigma_x)/2-\widehat{A_x}]\widehat{\Psi_{+}}=0$ with
$$
\widehat{A_x}=\frac{i}{4}\left(\begin{array}{cc}
2i(\partial_t-\partial_x)\varphi & m\lambda^{-1} (1-e^{2\varphi}) \\ 
 m\lambda^{-1} (1-e^{-2\varphi}) & -2i(\partial_t-\partial_x)\varphi
\end{array} \right)\,.
$$
The boundary conditions on $\varphi$ ensure that $\widehat{A_x}$ vanishes rapidly as $x\rightarrow \pm\infty$. This differential equation may be rewritten as an integral equation:
\begin{equation}
\widehat{\Psi_{+}}(x,\lambda)=\left(\begin{array}{c}1 \\ 1\end{array} \right)-\int_{x}^{\infty}
\dd y\, e^{\frac{i p_{\lambda}}{2} (y-x)(1-\sigma_x)}\widehat{A_x}(y,\lambda)\widehat{\Psi_{+}}(y,\lambda)\,.
\end{equation}
Decomposing this function into the two eigenvectors of $(1-\sigma_x)$,
$$
\widehat{\Psi_{+}}(x,\lambda)=F(x,\lambda)\left(\begin{array}{c}1 \\ -1\end{array} \right)
+G(x,\lambda)\left(\begin{array}{c}1 \\ 1\end{array} \right)\,,
$$
yields the coupled integral equations:
\begin{align}
F(x,\lambda)&=\int_{x}^{\infty}
\dd y\, e^{ip_\lambda(y-x)}\Big[
\frac{(\partial_t-\partial_x)\varphi}{2}G(y,\lambda)
\\&\qquad+\frac{im}{4\lambda}\Big(\left(1-\cosh(2\varphi)\right)F(y,\lambda)-\sinh(2\varphi)G(y,\lambda)
\Big)\Big]\\
G(x,\lambda)&=1+\int_{x}^{\infty}
\dd y\, e^{ip_\lambda(y-x)}\Big[
\frac{(\partial_t-\partial_x)\varphi}{2}F(y,\lambda)\\
&\qquad\qquad-\frac{im}{4\lambda}\Big(\left(1-\cosh(2\varphi)\right)G(y,\lambda)-\sinh(2\varphi)F(y,\lambda)
\Big)\Big]\,.
\end{align}
Note that for $f(x)$ decreasing sufficiently quickly as $x\rightarrow\infty$
$$
\int_x^\infty \dd y\, e^{i\lambda (y-x)}f(y)=O\left(\lambda^{-1}\right)\,,
$$
so that $G(x,\lambda)=1+O(\lambda^{-1})$ and $F(x,\lambda)=O(\lambda^{-1})$, as in equation~(\ref{EqLambAsym}). Derivations of the other three equations follow similarly.
%%%%%%%%%%%%%%%%%%%%%%%%%%%%%%%%%%%%%%%%%%%%%%%%%%%%%%%%%%
%%%%%%%%%%%%%%%%%%%%%%%%%%%%%%%%%%%%%%%%%%%%%%%%%%%%%%%%%%
\section{Deriving the large-$t$ expansion of the solution to the GLM equations} \label{appintegrals}
%%%%%%%%%%%%%%%%%%%%%%%%%%%%%%%%%%%%%%%%%%%%%%%%%%%%%%%%%%
%%%%%%%%%%%%%%%%%%%%%%%%%%%%%%%%%%%%%%%%%%%%%%%%%%%%%%%%%%
In order to show that the asymptotic form of the solution to (\ref{eqGLM1time}) is, to some extent, unique, it is useful to derive it from a more general trial solution. Let us consider (\ref{eqGLM1time}) without the integral over $F_{-1} \widehat{W_{1}}$. The pole as $x\to2t$ of the function $F_0^P(x,t)$ makes solving the equation (\ref{eqGLM1time}) quite subtle: this pole may give a pole to $\xi(x,y,t)$ as $y\to x$, which makes the integrals logarithmically divergent at the boundary $z=x$. Trying to take $\xi(x,y,t)$ as a principal part with respect to $x=y$ means that this divergency would be translated into some term $\log(\ep)$ -- the integral would not be a true principal value integral, and $\xi(x,y,t)$ not a true distribution. Due to the term with $\Theta(2t-x-y)$, there is also a possible problem at $y=2t-x$ in $\xi(x,y,t)$, where there could be a finite jump (in addition to a pole).  To make everything clear, we need to regularise the GLM equations. One could take from the beginning (\ref{presc}) with finite $\ep$ for the integrals (\ref{EqFInt}), or one could keep the number $\ep$ of the principal part prescription for $F^P_0(x,t)$ finite in the integral equations. For our purposes, it will be clearer to modify the integral over $z$ in the GLM equation (\ref{eqGLM1time}) by requiring $z-x>\varep$ and $|z-(2t-x)|>\varep$ for some fixed positive number $\varep$. Then, we may see the solution as a function of $\varep$, and the integral equations are true principal value integrals in the integration regions that are not excluded. We seek a solution that gives finite values for the quantities associated to the PDE, $\widehat{\Psi_\pm}$ and $\varphi$, as $\varep\to0$; we will find a solution that is itself finite as $\varep\to0$.

It will be convenient to replace the variables $x,y$ with $\Delta\equiv-x+t$ and $s\equiv y-t$. Our large $t$ region corresponds to $\Delta,m^{-1},T^{-1}\ll t$. Also, from the discussion above, the solution $\xi$ may separate into two regions (between which there could be a discontinuity):
\begin{itemize}
\item ``spacelike'', $\xi^s(\Delta,s,t)$: $s>\Delta$
\item ``timelike'', $\xi^t(\Delta,s,t)$: $-\Delta<s<\Delta$\,.
\end{itemize}
Discarding for a moment the contribution from the integral over $F_{-1}\widehat{W_1}$ (since it will be shown to contribute subleading terms), in terms of these variables, the GLM equation becomes
\begin{align}
-\frac{2}{m}&\sigma_z\xi^t(\Delta,s,t)= F_0^P(2t+s-\Delta,t)\left(\begin{array}{c}1 \\ 1\end{array}\right)+
F_0^P(2t+s+\Delta,t)\left(\begin{array}{c}-i \\ i\end{array}\right) \nonumber\\
&
+\int_{-\Delta+\varep}^{\Delta-\varep}{F_0^P(2t+s+s',t)\xi^t(\Delta,s',t)\dd s'} +\int_{\Delta+\varep}^{\infty}{F_0^P(2t+s+s',t)\xi^s(\Delta,s',t)\dd s'}
\nonumber\\
&+ \frac{2i}{m}\xi^t(\Delta,-s,t) 
\end{align}
and
\begin{align}
-\frac{2}{m}&\sigma_z\xi^s(\Delta,s,t)= F_0^P(2t+s-\Delta,t)\left(\begin{array}{c}1 \\ 1\end{array}\right)+
F_0^P(2t+s+\Delta,t)\left(\begin{array}{c}-i \\ i\end{array}\right) \nonumber\\
&
+\int_{-\Delta+\varep}^{\Delta-\varep}{F_0^P(2t+s+s',t)\xi^t(\Delta,s',t)\dd s'} +\int_{\Delta+\varep}^{\infty}{F_0^P(2t+s+s',t)\xi^s(\Delta,s',t)\dd s'} \;.
\end{align}
A linear combination of these equations and the same equations with $s\rightarrow -s$ yields the coupled equations for $\xi^s$ and $\xi^t$:
\begin{align}
&\frac{\xi^t(\Delta,s,t)}{m}=-\frac{\sigma_z}{4}\int_{-\Delta+\varep}^{\Delta-\varep}{F_0^P(2t+s+s',t)\xi^t(\Delta,s',t)\dd s'}
-\frac{\sigma_z}{4}\int_{\Delta+\varep}^{\infty}{F_0^P(2t+s+s',t)\xi^s(\Delta,s',t)\dd s'}\nonumber\\
& +\frac{i}{4}\int_{-\Delta+\varep}^{\Delta-\varep}{F_0^P(2t-s+s',t)\xi^t(\Delta,s',t)\dd s'}
+\frac{i}{4}\int_{\Delta+\varep}^{\infty}{F_0^P(2t-s+s',t)\xi^s(\Delta,s',t)\dd s'}\nonumber\\
&+\frac{i}{4} F_0^P(2t+s+\Delta,t)\left(\begin{array}{c}1 \\ 1\end{array}\right)
+\frac{\sigma_z}{4} F_0^P(2t-s+\Delta,t)\left(\begin{array}{c}1 \\ 1\end{array}\right)\nonumber\\
&-\frac{\sigma_z}{4} F_0^P(2t+s-\Delta,t)\left(\begin{array}{c}1 \\ 1\end{array}\right)
+\frac{i}{4} F_0^P(2t-s-\Delta,t)\left(\begin{array}{c}1 \\ 1\end{array}\right)\label{EqXit}\,,
\end{align}
and
\begin{align}
&\frac{\xi^s(\Delta,s,t)}{m}=-\frac{\sigma_z}{2}\int_{-\Delta+\varep}^{\Delta-\varep}{F_0^P(2t+s+s',t)\xi^t(\Delta,s',t)\dd s'}
-\frac{\sigma_z}{2}\int_{\Delta+\varep}^{\infty}{F_0^P(2t+s+s',t)\xi^s(\Delta,s',t)\dd s'}\nonumber\\
& +\frac{i}{2} F_0^P(2t+s+\Delta,t)\left(\begin{array}{c}1 \\ 1\end{array}\right)
-\frac{\sigma_z}{2} F_0^P(2t+s-\Delta,t)\left(\begin{array}{c}1 \\ 1\end{array}\right)
\label{EqXis}\,.
\end{align}

We are looking for the leading terms of the solution to these equations. More precisely, we will concentrate on terms which, in the limit $t\to\infty$ with $s,\Delta$ fixed, behave like $e^{-im\sqrt{2t}\sqrt{|s\pm\Delta|}}$ and $e^{-m\sqrt{2t}\sqrt{|s\pm\Delta|}}$, times powers of $t$. For instance, choosing $s=0$, this gives a precise large-time frequency of oscillation or exponential decay, and products of such terms give higher frequencies or stronger exponential decays. Terms with higher exponential decay or higher frequency will be deemed sub-leading. Hence, we will neglect all terms with $\nu\neq0$ in (\ref{expFj}) (and only the first series, with $\nu=0$, remains). It is these leading terms that will give the leading terms in our region of interest for the function $\varphi$ itself. With that understanding, for ease of notation, let us define the following four functions:
\begin{align}
q(\Delta,s,t)&\equiv F_0^P(2t+s+\Delta,t)\nonumber\\
u(\Delta,s,t)&\equiv F_0^P(2t-s+\Delta,t)\nonumber\\
v(\Delta,s,t)&\equiv F_0^P(2t+s-\Delta,t)\nonumber\\
w(\Delta,s,t)&\equiv F_0^P(2t-s-\Delta,t)~.
\end{align}
These are the inhomogeneous parts of the GLM equations for $\xi^t$ (equation~\ref{EqXit}) and $\xi^s$ (equation~\ref{EqXis}). Let us consider the trial solutions
\beq
	\xi^t=(\alpha q+\beta u+\gamma v+\delta w)\mato{c} 1\\ 1 \matf~,\quad
	\xi^s=(\alpha' q+\beta' u+\gamma' v+\delta' w)\mato{c} 1\\ 1 \matf~,
\eeq
where the coefficients are two by two matrices in $\R + \sigma_z \R$, and insert them into (\ref{EqXit}) and (\ref{EqXis}). It turns out that the leading terms (in the sense above) resulting from the integrals are of similar functional form, namely $t$- and $\varep$-dependent linear combinations of $q,u,v,w$ and of their $s$-derivatives to arbitrary high orders. Some details of the calculations to obtain the leading terms are done below. In fact, the leading power of $t^0$ in these linear combinations does not contain any derivative, so that to leading order these trial solutions give an inhomogeneous linear matrix equation for the coefficients. In matrix form, this equation is, to $O(\varep)$,
$$
M=\left(\begin{matrix}
\frac{1}{2} & -i\sigma_z-\frac{\sigma_z \t{K}}{4} &
-\frac{i\sigma_z}2 - \frc{\sigma_z K}{4} & 
0 & 0 & \frac{\sigma_z \t{K}}4 & \frac{\sigma_z K}4  & 0 \\ 
-\frac{i\sigma_z}{2} & \frac{i \t{K}}{4} & -\frc12 + \frc{i K}4 & 0 & 
0 & -\frac{i \t{K}}{4} & -\frac{i K}{4} & 0 \\ 
\frc{\sigma_z K}4 & 0 & \frac{1}{2} & -\frc{i\sigma_z}2 + \frc{\sigma_z \t{K}}4 & 
0 & 0 & 0 & 0 \\ 
-\frac{iK}{4} & 0 & -\frac{i\sigma_z}{2} & \frc12 - \frc{i\t{K}}4 & 
0 & 0 & 0 & 0 \\ 
0 & -i\sigma_z - \frc{\sigma_z\t{K}}2 & -i\sigma_z - \frc{\sigma_z K}2  & 0 & 1 & 
\frac{\sigma_z \t{K}}{2} & \frac{\sigma_z K}{2} & 0 \\ 
0 & 0 & 0 & 0 & 
0 & 1 & 0 & 0 \\ 
\frac{\sigma_z K}{2} & 0 & 0 & \frc{\sigma_z \t{K}}2 & 
0 & 0 & 1 & 0 \\ 
0 & 0 & 0 & 0 & 0 & 0 & 0 & 1
\end{matrix}\right)
$$
$$
M \left(\begin{matrix}
\alpha \\ \beta \\ \gamma \\ \delta \\ \alpha' \\ \beta' \\ 
\gamma' \\ \delta'\end{matrix} \right)=
m\left(\begin{matrix}
i/4 \\ \sigma_z/4 \\ -\sigma_z/4 \\ i/4 \\ 
i/2 \\ 0 \\ -\sigma_z/2 \\ 0 \end{matrix} \right)~,
$$
in which $K+\t{K}=-2i$ and $\pi K=4\gamma+4\ln(m\sqrt{\varep t}/2)$, with $\gamma$ Euler's constant. The linear combinations to all orders in negative powers of $t$ are formally reproduced by this matrix if $K$ and $\t{K}$ are, instead, power series in $t^{-1}$ whose coefficients are differential operators in $s$ of order equal to the power of $t^{-1}$ (and then, $K+\t{K}\neq-2i$). The matrix above is invertible for arbitrary $K$ and $\t{K}$ and has the miraculously simple solution:
\beq
\xi^t=-\frac{m\sigma_z}{2}v\left(\begin{array}{c}1 \\ 1\end{array}\right) ~,\quad
\xi^s=-\frac{m\sigma_z}{2}v\left(\begin{array}{c}1 \\ 1\end{array}\right)~.
\eeq
Since this is independent of $K$ and $\t{K}$, it is correct to $O(t^{-\infty})$. This gives the solution written in the text, (\ref{solxi}).

In order to obtain the matrix above, we use the expansion (\ref{expFj}) for the function $F_0^P(x,t)$ (the part with $\nu=0$ only). The idea is that if the large-$t$ limit (with $\Delta$ fix) can be taken in the integrands of (\ref{EqXit}), (\ref{EqXis}), then the result is sub-leading. The only problem in taking this limit is at the points in the integration region where the coefficient of $t$ in the argument of the Bessel functions in (\ref{expFj}) vanishes. This happens only for $2x=t$ in $F_0^P(x,t)$ (that is, at its singular point -- this is why the regularisation $\varepsilon$ is important). There are two types of points where this happens. First, the kernels $F_0^P(2t\pm s+s')$ of the integral equations (\ref{EqXit}), (\ref{EqXis}) give a principal-value integral about the point $\pm s + s'=0$. The $s'$-contour can completed and deformed towards the positive imaginary direction (this direction is chosen due to the square-root structure of the argument of the Bessel functions), and the half-residue substracted. The resulting half-residues give contributions to the leading terms of interest.

The second type of points are those coming not from the kernels, but from the functions $q,u,v,w$ in the trial solution. For the functions $q$ and $w$, they are at the boundary $s'=-\Delta+\varep$; for $u$ and $v$, they are at $\Delta\pm\varep$. Let us give an example with the function $q$. We have, for arbitrary $g(s')$,
\beqa
	\int_{-\Delta+\varep} g(s') q(\Delta,s',t) \dd s' &=& \int_{-\Delta+\varep} g(s') F_0^P(2t+s'+\Delta) \dd s' \n
		&\sim& \frc{i}\pi \sum_{\mu=0}^\infty \int_{\varep} g(s'-\Delta) \lt(\frc{-i s'}{i(4t+s')}\rt)^{\frc{\mu-1}2} K_{1-\mu}(m\sqrt{i(2t+s'/2)(-is'/2)}) \dd s' \n
	&=& \frc{2i}{\pi t} \sum_{\mu=0}^\infty \int_{\sqrt{t\varep}} g(l^2/t-\Delta) e^{\frc{i\pi(1-\mu)}2}\lt(\frc{l^2}{4t^2+l^2}\rt)^{\frc{\mu-1}2}
		K_{1-\mu}(ml\sqrt{1+l^2/(4t^2)}) l \dd l \no
\eeqa
With the change of variable in the last line, the point $s'=-\Delta$ has been ``resolved'', and the large-$t$ expansion can be taken in the integrand. This gives
\[
	m^{-1} (K g)(-\Delta)
\]
where $K$ is an expansion in inverse powers of $t$ with coefficients that are, generically, differential operators. More precisely,
\[
	K = 4\gamma+4\ln(m\sqrt{\varep t}/2) + \frc1{mt} \lt(\frc{2i c_{10}}{\pi} - \frc{8}{\pi m} \partial_s\rt) + \ldots
\]
where $s$ is the independent variable of the function on which it applies. Similarly,
\[
	\int_{-\Delta+\varep} g(s') w(\Delta,s',t) \dd s' \sim m^{-1} (\t{K}g)(-\Delta)
\]
where $\t{K}$ is a different operator. Other integrals give
\beqa
	\int^{\Delta-\varep} g(s') u(\Delta,s',t) \dd s' &\sim& m^{-1} ( (-2i-\t{K})g)(\Delta) \n
	\int^{\Delta-\varep} g(s') v(\Delta,s',t) \dd s' &\sim& m^{-1} ( (-2i-K)g)(\Delta) \n
	\int_{\Delta+\varep} g(s') u(\Delta,s',t) \dd s' &\sim& m^{-1} ( \t{K}g)(\Delta) \n
	\int_{\Delta+\varep} g(s') v(\Delta,s',t) \dd s' &\sim& m^{-1} ( Kg)(\Delta)~.\no
\eeqa
Using these, one gets the other terms giving the 8 by 8 matrix above.

%%%%%%%%%%%%%%%%%%%%%%%%%%%%%%%%%%%%%%%%%%%%%%%%%%%%%%%%%%
%%%%%%%%%%%%%%%%%%%%%%%%%%%%%%%%%%%%%%%%%%%%%%%%%%%%%%%%%%

%%%%%%%%%%%%%%%%%%%%%%%%%%%%%%%%%%%%%%%%%%%%%%%%%%%%%%%%%%
%%%%%%%%%%%%%%%%%%%%%%%%%%%%%%%%%%%%%%%%%%%%%%%%%%%%%%%%%%
\end{document}